\documentclass[preprint,3p,times]{elsarticle}

\usepackage{lineno}

\usepackage{
    amsmath,
    amssymb,
    amsthm,
    MnSymbol, %
    mathbbol,
    amsfonts,
    ifthen,
}

\newtheorem{definition}{Definition}[section]

\newtheorem{proposition}[definition]{Proposition}
\newtheorem{theorem}[definition]{Theorem}
\newtheorem{corollary}[definition]{Corollary}

\newtheorem{example}[definition]{Example}

\newtheorem{remark}[definition]{Remark}

\usepackage[utf8]{inputenc}                 %
\usepackage{color}

\usepackage{quiver}

\usepackage{hyperref}
\hypersetup{hidelinks} %

\newcommand{\Functor}[8][cccc]{
    \ensuremath{
        \ifthenelse{\equal{#2}{}}{}{#2 : \left\{ }
        \begin{array}{#1}
            #3  & \longrightarrow   & #4 \\
            #5  & \longmapsto       & #6 \\
            #7  & \longmapsto       & #8
        \end{array}
        \ifthenelse{\equal{#2}{}}{}{\right.}
    }
}

\newcommand{\Function}[6][\longrightarrow]
{%
	\ensuremath{%
		\ifthenelse{\equal{#2}{}}{}{#2 : \left\{ }%
		\begin{array}{ccc}%
			#3 	&	 #1           & 	#4 		\\%
			#5 	& 	\longmapsto   & 	#6%
		\end{array}%
		\ifthenelse{\equal{#2}{}}{}{\right.}%
	}%
}

\newcommand{\Set}{\mathrm{Set}}

\newcommand{\Sub}{\mathrm{Sub}}
\newcommand{\Hom}{\mathrm{Hom}}
\newcommand{\nat}{\mathrm{Nat}}
\newcommand{\Sieve}{\mathrm{Sieve}}

\newcommand{\dom}{\mathrm{dom}}
\newcommand{\codom}{\mathrm{cod}}
\newcommand{\im}{\mathrm{Im}}

\newcommand\Omit[1]{}

\def\sem#1{[\![ #1 ] \!]}

\newcommand{\Ctx}{\mathcal{C}tx}
\newcommand{\Qf}{\mathcal{Q}f}
\newcommand{\Heytalg}{HeytAlg}
\newcommand{\Los}{\L{}o{\'s}}
\newcommand{\mto}{\rightarrowtail}
\renewcommand{\hat}{\widehat}
\newcommand{\set}[1]{\{#1\}}                                %
\newcommand{\bigset}[1]{\big\{#1\big\}}                     %

\makeatletter
\def\bstr{b}
\def\bfstr{bf}
\def\cstr{c}
\def\fstr{f}
\def\sstr{s}

\def\strLst{A,B,C,D,d,E,F,G,H,I,J,K,L,M,N,O,P,Q,R,S,T,U,V,W,X,Y,Z}

\newcommand{\MkB}[1]{\expandafter\def\csname\bstr#1\endcsname{\mathbb{#1}}}
  \@for\i:=\strLst\do{%
    \expandafter\MkB \i     }
    
\newcommand{\MkBF}[1]{\expandafter\def\csname\bfstr#1\endcsname{\mathbf{#1}}}
  \@for\i:=\strLst\do{%
    \expandafter\MkBF \i     }

\newcommand{\MkCal}[1]{\expandafter\def\csname\cstr#1\endcsname{\mathcal{#1}}}
  \@for\i:=\strLst\do{%
    \expandafter\MkCal \i     }
    
\newcommand{\MkFrak}[1]{\expandafter\def\csname\fstr#1\endcsname{\mathfrak{#1}}}
  \@for\i:=\strLst\do{%
    \expandafter\MkFrak \i     }
    
\newcommand{\MkSF}[1]{\expandafter\def\csname\sstr#1\endcsname{\mathsf{#1}}}
  \@for\i:=\strLst\do{%
    \expandafter\MkSF \i     }
    
\makeatother %

\journal{Annals of Pure and Applied Logic}
\newpageafter{abstract}

\begin{document}

\begin{frontmatter}

\title{Ultraproducts in abstract categorical logic}

\author[mics]{Marc Aiguier}
\ead{marc.aiguier@centralesupelec.fr}
\author[lip6]{Isabelle Bloch}
\ead{isabelle.bloch@sorbonne-universite.fr}
\author[mics,kit]{Romain Pascual}
\ead{romain.pascual@kit.edu}
\affiliation[mics]{%
	organization={MICS, CentraleSupelec, Universite Paris Saclay},
	country={France}
}
\affiliation[lip6]{%
	organization={Sorbonne Universite, CNRS, LIP6},
	city={Paris},
	postcode={F-75005},
	country={France}
}
\affiliation[kit]{%
	organization={Karlsruhe Institute of Technology},
	city={Karlsruhe},
	country={Germany}
}

\begin{abstract}
In a previous publication, we introduced an abstract logic via an abstract notion of quantifier. Drawing upon concepts from categorical logic, this abstract logic interprets formulas from context as subobjects in a specific category, e.g., Cartesian, regular, or coherent categories, Grothendieck, or elementary toposes. We proposed an entailment system formulated as a sequent calculus which we proved complete. Building on this foundation, our current work explores model theory within abstract logic. More precisely, we generalize one of the most important and powerful classical model theory methods, namely the ultraproduct method, and show its fundamental theorem, i.e., \Los's theorem. The result is shown as independently as possible of a given quantifier.
\end{abstract}

\end{frontmatter}

\tableofcontents

\section{Introduction}

The need to abstract the notion of logic emerged as a response to the profusion of logic in mathematics and computer science. In the 1930s, A. Tarski and his Polish school proposed a generalization of the inference relation $\vdash$~\cite{Tar56}. 
This generalization was only of syntactic nature, leaving aside any consideration for the semantics. The absence of an abstract notion of semantics prevented the generalization of any result from model theory. Among the constructions of abstract logic with a generalized notion of semantics, we can cite J. Barwise's approach~\cite{Bar74} and the theory of institutions~\cite{GB92}. Although being the first work generalizing semantics, Barwise's construction only dealt with extensions of first-order logic (FOL). Some model theory results could be established within this framework, the most famous being Lindstrom's theorem~\cite{Lin69}, which characterizes FOL in terms of fundamental theoretic properties (compactness and Loweineim-Skolem theorem). Institutions provided a much deeper generalization, addressing software specification and semantics issues. The extension provided by institutions is manifold:

\begin{itemize}
	\item institutions include a notion of signature category;
	\item sentences are members of a set (syntax free, as in Tarskian logics) and models are objects of a category (semantics free);
	\item institutions preserve the renaming property extended to any signature morphism. This property is called the satisfaction condition.
\end{itemize}

The theory of institutions provided a framework for generalizing many results across computer science~\cite{MA:AIJ-18,MA:IJAR-18,AB19,Bor02,Tar99} and model theory~\cite{AD07,Dia08,GP10,GP07}. In contrast to Barwise's approach, institutions eschew any presumption regarding the internal structure, neither for the sentences nor for the models (although it remains feasible to define propositional connectives and FOL quantifiers internally~\cite{Dia08}). However, generalizing standard results often requires the closure of formulas under some or all propositional connectives and FOL quantifiers. For instance, Robinson's consistency~\cite{GP07} and Craig's interpolation~\cite{Dia04} in institutions suppose the closure of sentences under negations, finite conjunctions, and universal quantification. Similarly, the formalization of abductive reasoning of~\cite{MA:IJAR-18} within institutions assumes sentences closed under propositional connectives. In~\cite{AB23}, we proposed to define abstractly the notion of logic by supposing that, 

\begin{enumerate}
    \item as in institutions, no commitment is made to the internal structure of models except that they have a carrier taken in a category with some properties (called prop-category in the paper -- see Definition~\ref{def:prop-category} below),
    \item unlike in institions, formulas are inductively defined from propositional connectives, an abstraction of atomic formulas (called basic formulas in the paper), and an abstract notion of quantifiers.
\end{enumerate}

As in categorical logic, we defined in~\cite{AB23} semantics by interpreting formulas from context as subobjects of an object of a given category (Cartesian\footnote{Also called cartesian monoidal categories.}, regular, coherent, Grothendieck toposes, elementary toposes -- see~\cite{Johnstone02}).
Therefore, the semantic framework of~\cite{AB23} abstracts both contexts and subobjects in the spirit of Lawvere's hyperdoctrines~\cite{Law69,Law70}. Subsequently, our abstract logic follows the principles of categorical logic: an internal logic has been defined as an extension of propositional logic (PL) over this semantic framework.

\medskip
In this abstract logic, as is customary in categorical logic, we proposed an entailment system formulated as a sequent calculus for which we proved a completeness result. Here, we propose to pursue the generalization of model theory results. More precisely, we propose to generalize one of the most important and powerful classical model theory methods, namely the ultraproducts method~\cite{CK73}. Hence, we propose to study conditions for the development of the ultraproducts method as independently as possible of quantifiers.

\medskip
Ultraproducts are quotients of directed products of a family of structures, typically used in abstract algebra and logic, especially in model theory. The authors in~\cite{bergelson_ultrafilters_2010} provide various applications of ultraproducts to model theory, algebra, and nonstandard analysis. In particular, the chapter by J. Keisler~\cite{keisler_ultraproduct_2010} surveys the classical results on ultraproducts of first-order structures. J. Keisler explains that the idea goes back to the construction of nonstandard models of arithmetic by T. Skolem~\cite{skolem_uber_1934}, then studied for fields by E. Hewitt~\cite{hewitt_rings_1948} before a generalization to first-order structures by J. \Los~\cite{los_quelques_1955}. S. Galbor also surveys applications where methods based on ultrafilters play significant roles~\cite{sagi_ultraproducts_2023}: in model theory for the compactness theorem, theorems about axiomatizability, and characterizations of elementary equivalence, in algebra to construct new fields such as the hyperreal numbers, or in nonstandard analysis for the theories of infinitesimal numbers.

\medskip
The ultraproducts method has already been abstractly investigated within the framework of FOL and its restrictions~\cite{AN78,Dia08,Dia17}. We will show that some of these results can be considered as an instance of our main result. 
In the context of categorical logic, a more general result than \Los's theorem (in the sense that it implies the classical version in \(\Set\)) has been demonstrated by replacing the notion of first-order theory with that of pretopos and by defining models as functors from a pretopos to Set~\cite{Mak87}. Classical \Los's theorem is then derived by replacing pretopos with the syntactic category of the suitable first-order theory.
In~\cite{Mak87}, M. Makkai also showed that any small pretopos $\cC$ can be recovered from the category of models $Mod(\cC)$ together with some additional structure given by the ultraproduct construction (ultracategories).

\medskip
Note that no assumption is made on the category of models in our abstract categorical logic (except, of course, small products on which filtered products are defined). This lack of structure associated with the interpretation of formulas is the source of the main difficulty in proving \Los's result. Indeed, the standard notion of filtered products is built componentwise on models and only provides finite intersections. Consequently, more information is required to prove the result in the case of basic formulas (also called atomic formulas). Intuitively, the base case of the induction proof of \Los's result can only be obtained by assuming some conditions. Hence, we will provide necessary conditions -- namely a sup-generation condition and a finiteness condition -- that guarantee the result for basic formulas. These conditions echo those given by R. Diaconescu for institutions~\cite{Dia08}.

The main contributions of this paper are a generalization of the ultraproduct method and a generalization of \Los's theorem, the fundamental result on ultraproducts, both independently of quantifiers. Additionally, the relevance of all introduced definitions and results is demonstrated on several different logics throughout the paper.

\medskip
In the preliminary section, besides briefly reviewing some terminology, concepts, and notations about filters and filtered colimits, we recall the categorical definitions of filtered products and present the notion of prop-categories, which generalizes the standard notion of subobjects in category theory. %
Section~\ref{sec:abstract categorical logic} reviews the abstract categorical logic defined in~\cite{AB23} via the notions of semantical systems, quantifiers, and internal logic. Section~\ref{sec:ultraproduct} extends the fundamental theorem of ultraproducts as independently as possible of a given quantifier. Note that Section~\ref{sec:abstract categorical logic} is substantially similar to Sections~2,~3, and~4 of~\cite{AB23}. However, we add additional properties to the various notions, which we then leverage to obtain \Los's theorem in Section~\ref{sec:ultraproduct}, which is the main result presented in this paper.

\section{Preliminaries}

We assume familiarity with the main notions of category theory, such as categories, functors, natural transformations, limits, colimits, and Cartesian closedness, and refer the interested reader to textbooks such as~\cite{BW90,McL71}. We also assume basic knowledge of first-order and modal logics (ML)~\cite{CK73,Che80}.

\subsection{Notations}

In the whole paper, $\cC$ and $\cD$ denote generic categories, $X$ and $Y$ denote objects of categories (the collection of objects of a category $\cC$ is written $|\cC|$). When a category $\cC$ is Cartesian closed, $X^Y$ denotes the exponential object of $X$ and $Y$. The symbols $f$, $g$, and $h$ denote morphisms, and given a morphism $f : X \to Y$, we write $\dom(f) = X$ for the domain of $f$ and $\codom(f) = Y$ for its co-domain; $F,G,H : \cC \to \cD$ denote functors from a category $\cC$ to a category $\cD$, $F^{op}$ the opposite functor of $F$, $\alpha,\beta : F \Rightarrow G$ natural transformations, and $Nat(F;G)$ the class of all natural transformations between $F$ and $G$. Identity morphisms are written $Id$, and initial and terminal objects $\emptyset$ and $\mathbb{1}$, respectively. Finally, monomorphisms are denoted by $\mto$, i.e., if $m$ is a monomorphism from $X$ into $Y$, then we write $m: X \mto Y$.

\subsection{Prop-categories}
\begin{definition}[Prop-category]
\label{def:prop-category}
A {\bf prop-category} $\cC$ is a category %
with a contravariant functor $Prop_\cC : \cC^{op} \to \Heytalg$ where $\Heytalg$ is the category of Heyting algebras.
Given an object $X \in |\cC|$, the lower and upper bounds of $Prop_\cC(X)$ are respectively denoted by $\bot_X$ and $\top_X$, its order by $\preceq_X$, or simply $\preceq$ when there is no ambiguity, 
and the meet, join and implication operations respectively by $\wedge$, $\vee$, and $\rightarrow$.\footnote{In a Heyting algebra, the implication $\to$ is right-adjoint to the meet operation $\wedge$, i.e., given a Heyting algebra \((\cH,\preceq_\cH)\), for all elements \(a\) and \(b\) in \(\cH\), there exists a greatest element \(c\) in \(\cH\) such that \(a \wedge c \preceq_H b\), which is \(a \rightarrow b\).
}

Given a morphism $f : X \to Y \in \cC$, $Prop_\cC(f) : Prop_\cC(Y) \to Prop_\cC(X)$ is called the {\bf pullback functor} or {\bf base change} along $f$ (the posets $Prop_\cC(X)$ and $Prop_\cC(Y)$ are considered as categories). It will be denoted $f^*$, i.e., $f^* = Prop_\cC(f)$. 
\end{definition}

\begin{remark}
    Compared to~\cite{AB23}, we now require that the functor \(Prop_\cC\) is a functor \(\cC^{op} \to \Heytalg\) rather than \(\cC^{op} \to Pos\) where \(Pos\) is the category of posets. Indeed, Heyting algebras allow for the interpretation of propositional logic (PL) connectives. %
\end{remark}
 
Prop-categories generalize the notion of subobjects in an arbitrary category. We recall the categorical notion of subobjects to explain the generalization. 
Given an object \(X\) in a category \(\cC\), the set of monomorphisms into \(X\) admits a preorder \(\preceq_X\) such that \(a: A \mto X\) is less than or equal to \(b: B \mto X\) (i.e., $a \preceq_X b$) whenever there exists a morphism \(x: A \to B\) such that \(a = b \circ x\).

Then, a subobject of \(X\) is an equivalence class for the equivalence relation \(\simeq_X\) induced by \(\preceq_X\), and $\Sub(X)$ is the set of equivalence classes for $\simeq_X$. The preorder \(\preceq_X\) yields a partial order on \(\Sub(X)\), which we also write \(\preceq_X\). We will also identify the equivalence classes of \(\Sub(X)\) with any of its representatives. For instance, in \(\Set\), the category of sets and functions, subobjects are subsets, while in \(Graph\), the category of graphs and graph morphisms, subobjects are subgraphs.

When \(\cC\) has pullbacks, subobjects give rise to the contravariant functor $\Sub : \cC^{op} \to Pos$ which to every $X \in |\cC|$ associates $\Sub(X)$ and to every morphism $f : X \to X'$ associates the mapping $\Sub(f) : \Sub(X') \to \Sub(X)$ which to every $Y' \mto X'$ associates $Y \mto X$ making the diagram
\[\begin{tikzcd}
	Y & {Y'} \\
	X & {X'}
	\arrow[from=1-1, to=1-2]
	\arrow[tail, from=1-2, to=2-2]
	\arrow[tail, from=1-1, to=2-1]
	\arrow[from=2-1, to=2-2]
\end{tikzcd}\]
a pullback.

To equip $\Sub(X)$ with a structure of Heyting algebra, the category $\cC$ must satisfy additional properties fulfilled, for instance, by elementary toposes. Elementary toposes, together with the contravariant functor $\Sub$, are the archetypical class of examples of prop-categories~\cite{Johnstone02}, which we will use. We now recall the definition of elementary toposes alongside their main results.

\subsubsection{Elementary topos}
\label{sec:elementary topos}

An elementary topos $\cC$ is a finitely complete Cartesian closed category with a subobject classifier $\Omega$. Having a subobject classifier means that there is a morphism out of the terminal object $true : \mathbb{1} \to \Omega$ such that for every monomorphism $m : Y \mto X$ there is a unique morphism $\chi_m : X \to \Omega$ (called the characteristic morphism of $m$) such that the following diagram is a pullback:
\[\begin{tikzcd}
	Y & {\mathbb{1}} \\
	X & \Omega
	\arrow["{!}", from=1-1, to=1-2]
	\arrow["true", tail, from=1-2, to=2-2]
	\arrow["m"', tail, from=1-1, to=2-1]
	\arrow["{\chi_m}"', from=2-1, to=2-2]
\end{tikzcd}\]

When \(\cC\) is an elementary topos, $\Sub(X)$ is a Heyting algebra~\cite{Johnstone02}, and $(\Sub(X),\preceq_X)$ forms a distributive bounded lattice with $Id_X$ and $\emptyset \mto X$ as the largest and smallest elements, respectively, and which admits an implication $\to$ right-adjoint to the meet operation $\wedge$. Note that an elementary topos is also finitely cocomplete, i.e., has finite colimits, and therefore, it has an initial object $\emptyset$ which is the colimit of the empty diagram, meaning that $\emptyset \mto X$ is always well-defined.

The following properties hold in any topos~\cite{BW85,Johnstone02}:

\begin{itemize}
\item Every morphism $f$ can be factorized uniquely as $m_f \circ e_f$ where $e_f$ is an epimorphism and $m_f$ is a monomorphism. The codomain of $e_f$ is often denoted by $\im(f)$ and is called the {\em image of} $f$, and then $(A \stackrel{f}{\rightarrow} B) = (A \stackrel{e_f}{\rightarrow} \im(f) \stackrel{m_f}{\mto} B)$.

\item Every object $X \in |\cC|$ has a {\em power object} defined by $\Omega^X$ and denoted $PX$. As a power object, it satisfies the following adjunction property: 
$$\Hom_\cC(X \times Y,\Omega) \simeq \Hom_\cC(X,PY)$$

Given a morphism $f \in \Hom_\cC(X \times Y,\Omega)$ (respectively $f \in \Hom_\cC(X,PY)$) we denote by $f^\#$ its equivalent by the above bijection. The morphism $f^\#$ is called the {\em transpose} of $f$. Note that, by construction, we have $(f^\#)^\# = f$.

In particular, the transpose of the identity $Id_{PX} : PX \to PX$ is the characteristic morphism of a subobject $\ni_X \mto PX \times X$.

\end{itemize}

As in the category $\Set$, the power object function which maps every object $X \in |\cC|$ to its power object $PX$ can be extended into a contravariant functor $\cP : \cC \to \cC$ which associates to every morphism $f : X \to Y$ the morphism $\cP f : PY \to PX$ whose transpose classifies the morphism $R \mto PY \times X$ where $R$ is the pullback of the diagram
\[\begin{tikzcd}
	R && {\ni_Y} \\
	{PY \times X} && {PY \times Y}
	\arrow[from=1-1, to=1-3]
	\arrow[tail, from=1-3, to=2-3]
	\arrow[tail, from=1-1, to=2-1]
	\arrow["{Id_X \times f}"', from=2-1, to=2-3]
\end{tikzcd}\]
Likewise, the power object function can also be extended into a covariant functor $\exists$ which associates to every morphism $f : X \to Y$ the morphism $\exists f : PX \to PY$ whose transpose classifies the image of the morphism 
\[
g :\,\ni_X \mto PX \times X \stackrel{Id_{PX} \times f}{\to} PX \times Y,
\]
i.e., $\exists f = \chi_{Im(g) \mto PX \times Y}^\#$.

\medskip
By the bijection $Hom_\cC(\mathbb{1},PX) \simeq \Sub(X)$, the morphism $\exists f : PX \to PY$ gives rise to a Heyting morphism $\exists f : \Sub(X) \to \Sub(Y)$ satisfying (see~\cite{Johnstone02}):\footnote{It is also known that $f^*$ has a right-adjoint $\forall f$ which makes the functor $\Sub$ a tripos over $\cC$ (see \cite{Pit02} for a restropective on this subject).}
\[
    \exists f(A \mto X) \preceq_Y B \mto Y \; \mbox{iff} \; A \mto X \preceq_X f^*(B \mto Y).
\]

Toposes are sufficiently set-behaved to internalize a logic in which one may reason as if they were picking elements in a set and accommodate internally constructive proofs, i.e., using neither the law of excluded middle nor the axiom of choice.
This internal language of toposes is recalled in Appendix~\ref{sec:AppB}, and will be used in the remainder of this paper.

\medskip
There are a multitude of examples of toposes. The most emblematic is Grothendieck toposes, defined as any category equivalent to the category of sheaves over a site~\cite{Car17}. 
Interestingly, presheaves form a simpler case of toposes and subsume most algebraic structures used in computer science, such as sets, graphs, and hypergraphs.
We now detail the case of presheaves, both their construction and the fact that they form toposes.

\subsubsection{Presheaf categories \ensuremath{\widehat{\cC}} as a special example of toposes}
 
Let $\cC$ be a small category, i.e., both collections of objects and arrows are sets. Let us denote by $\widehat{\cC}$ the category of contravariant functors $F : \cC^{op} \to \Set$ (presheaves).\footnote{We use here the French notation $\widehat{\cC}$ to denote the category of presheaves over a base category $\cC$.} Since $\cC$ is a small category, $\widehat{\cC}$ is complete and co-complete (i.e., it has all limits and colimits). We now detail why it is also a topos (a different proof, relying on a different construction of toposes, may be consulted in~\cite[Section 2.1, Theorem 4]{BW85}). First of all, observe that the functor $\Sub : \widehat{\cC} \to \Set$, which maps every presheaf $F$ to its set of subobjects $\Sub(F)$, is naturally isomorphic to the functor which maps each presheaf $F$ to the set of its sub-presheaves. Therefore, we can assume that $G(C) \subseteq F(C)$ for all $G \in \Sub(F)$ and $C \in |\cC|$, and then such a subobject will be denoted by $G \subseteq F$.

\paragraph{$\widehat{\cC}$ is Cartesian closed.} The product of two functors $F ,G : \cC^{op} \to \Set$ is the functor $H : \cC^{op} \to \Set$ defined for every $C \in |\cC|$ by $H(C) = F(C) \times G(C)$, and for every $f : A \to B \in \cC$ by the mapping $H(f) : H(B) \to H(A)$ defined by {$(a,b) \mapsto (F(f)(a),G(f)(b))$}. 

By Yoneda's Lemma, the exponential of functors $F,G : \cC^{op} \to \Set$ to the object $C \in |\cC|$ should give an isomorphism $G^F(C) \simeq \nat(\Hom(\_,C),G^F)$. However, the definition of Cartesian closedness requires that 
\[
\nat(\Hom(\_,C),G^F) \simeq \nat(\Hom(\_,C) \times F,G).
\] 
This leads naturally to define the exponential of $F$ and $G$ by the functor $G^F$, which associates to any object $C \in |\cC|$, the set of natural transformations from $\Hom(\_,C) \times F$ to $G$. For every $f : A \to B \in \cC$, $G^F(f) : G^F(B) \to G^F(A)$ is the mapping which associates to any natural transformation $\alpha : \Hom(\_,B) \times F \Rightarrow G$ the natural transformation $\beta : \Hom(\_,A) \times F \Rightarrow G$ defined for every object $C \in |\cC|$ by $\beta_C(g : C \to A,c \in F(C)) = \alpha_C(f \circ g,c)$.

\paragraph{$\widehat{\cC}$ has a subobject classifier.} For every $A \in |\cC|$, a set $S$ of arrows $f$ in $\cC$ is said to be a \textbf{sieve on $A$} if $S$ is a set of morphisms with codomain $A$ closed under precomposition with morphisms in $\cC$, i.e.:

\begin{enumerate}
\item For all arrows $f \in S$ we have $\codom(f) = A$, and 
\item For all arrows $f \in S$ and $g \in {\Hom}(C)$ such that $\codom(g) = \dom(f)$, we have $f \circ g \in S$. 
\end{enumerate}

We write $\Sieve(A)$ for the set of sieves on $A$. Moreover, the map $\Sieve : \cC \to \Set$ is naturally extended to a \textit{contravariant} functor $\Omega : \cC \to \Set$, i.e., a presheaf $\Omega \in \widehat{\cC}$, as follows:
\[
\Functor{\Omega}{\cC}{\Set}{A}{\Sieve(A)}{f : A \to B}{\Function{\Omega(f)}{\Sieve(B)}{\Sieve(A)}{S}{\{g : C \to A \mid f \circ g \in S\}}}
\]

In fact, $\Omega$ is the subobject classifier. Indeed, let us consider:
\begin{itemize}
    \item the natural transformation $true : \mathbb{1} \Rightarrow \Omega$ which\footnote{$\mathbb{1} : \cC^{op} \to \Set$ is the presheaf which associates to any $A \in |\cC|$ the terminal object $\mathbb{1}$ in $\Set$.} for every $A \in \cC$ associates to the unique element in $\mathbb{1}(A)$ the maximal sieve on $A$ (i.e., the unique sieve which contains $Id_A$);
    \item for every presheaf $F \in |\widehat{\cC}|$, the bijection:
\[
\Function{\chi}{\Sub(F)}{\Hom(F,\Omega)}{G \subseteq F}{\Function{\chi(G)_A}{F(A)}{\Sieve(A)}{x}{\{f : B \to A \; |\; F(f)(x) \in G(B) \}} }
\]
whose inverse is:
\[
\Function{\chi^{-1}}{\Hom(F,\Omega)}{\Sub(F)}{\xi}{A \mapsto \{x \in F(A) \; |\; Id_A \in \xi(A)(x) \}}
\]
\end{itemize}

Then we clearly have a correspondence between subobjects of $F \in |\widehat{\cC}|$ and morphisms $F \to \Omega$, via the following pullback: 
\[\begin{tikzcd}
	G & {\mathbb{1}} \\
	F & \Omega
	\arrow["{!}", from=1-1, to=1-2]
	\arrow["true", tail, from=1-2, to=2-2]
	\arrow["i"', tail, from=1-1, to=2-1]
	\arrow["{\chi(G)}"', from=2-1, to=2-2]
\end{tikzcd}\]
This makes $\Omega$ a subobject classifier in $\widehat{\cC}$.
Hence, given a presheaf $X : \cC^{op} \to \Set$, the power object $PX : \cC^{op} \to \Set$ is the presheaf which, given an object $C \in |\cC|$, gives the set 
$$PX(C) = \nat(\Hom_\cC(-,C) \times X, \Omega) \simeq \Sub(\Hom_\cC(-,C) \times X)$$

To sum up this subsection, prop-categories build on a generalization of subobjects via the functor \(Prop\). The objects of a prop-category will serve as the carriers of models, while the functor \(Prop\) will permit distinguishing between the values that validate the formula and those that do not. Classic examples of prop-categories are elementary toposes, in particular, presheaf toposes, subsuming sets, and graph-like structures.

\subsection{Filters, filtered colimits}

\subsubsection{Set-theoretic filters and ultrafilters}

Let $I$ be a nonempty set. A {\bf filter} $F$ {\bf over} $I$ is a subset of $\powerset(I)$ such that:\footnote{$\powerset(I)$ denotes the powerset of $I$.}
\begin{itemize}
	\item $I \in F$;
	\item if $A,B \in F$, then $A \cap B \in F$, and
	\item if $A \in F$ and $A \subseteq B$, then $B \in F$.
\end{itemize}
A filter $F$ is {\bf proper} when $F$ is not $\powerset(I)$, and it is an {\bf ultrafilter} when for every $A \in \powerset(I)$, $A \in F$ if, and only if $I \setminus A \notin F$. In particular, this implies that if \(F\) is an ultrafilter, then $\emptyset \notin F$.

\medskip
Some examples of filters are:

\begin{itemize}
	\item The trivial filter on a set $I$ is $F = \{I\}$.
	\item The filter generated by some $J \subseteq I$ is $F_J = \{A \in \powerset(I) \mid J \subseteq A\}$. It is called a {\bf principal filter}. If $I$ is finite, all filters on $I$ are principal. 
	\item Assume $I$ is infinite. {\bf Fr\'echet's filter} is defined as 
	$$F_\infty = \{A \in \powerset(I) \mid A \; \mbox{is cofinite}\}$$ 
	This filter is not principal. Indeed, let $J \in F_\infty$ and let $i_0 \in J$. $J \setminus \{i_0\}$ is still cofinite. 
\end{itemize}
An ultrafilter is then a maximal filter for the inclusion. Using Zorn's lemma, it is easy to see that any filter is contained in an ultrafilter. 

A conventional approach to the satisfaction of a formula in a context is to interpret it as the set of values that validate it. Hyperdoctrines introduced by Lawvere~\cite{Law69,Law70} generalize this approach to categorical logic. Here, we follow Pitt's terminology to hyperdoctrines~\cite{Pit95}, as we did in~\cite{AB23}.

\subsubsection{Filtered products in categories}
\label{sec:filteredproducts}

The standard definition of ultraproducts (e.g., in FOL) relies on constructing filtered products where the filter is an ultrafilter. Therefore, the central construction is that of filtered products. The general concept of filtered products in arbitrary categories comes from the consideration that filtered products in FOL can be seen as directed colimits of products of models. To our knowledge, this generalization first appeared in~\cite{Okh66}. This definition of filtered products as colimits of directed diagrams of projections between the (direct) products determined by the corresponding filter is a particular instance of the categorical concept of a reduced product. It has become the de-facto construction~\cite{AN78,Dia08,Dia17}.

\begin{definition}[Filtered product]
Let \(F\) be a filter over a set of indices $I$, and let $X = (X_i)_{i \in I}$ be an $I$-indexed family of objects in $\cC$. Then, \(F\) and \(X\) induce a functor $A_F : F^{op} \to \cC$, mapping each subset inclusion $J \subseteq J'$ of $F$ to the canonical projection $p_{J',J}: \prod_{J'} X_i \to \prod_J X_i$.

The {\bf filtered product} of $X$ modulo \(F\) is the colimit $\mu : A_F \Rightarrow \prod_F X$ of the functor $A_F$.
Diagrammatically, this reads
\[\begin{tikzcd}
	{\Pi_{j\in J'} X_j} && {\Pi_{j \in J} X_j} \\
	& \textcolor{rgb,255:red,92;green,92;blue,214}{\Pi_F X} \\
	& \textcolor{rgb,255:red,214;green,92;blue,92}{Y}
	\arrow["{P_{J',J}}", color={rgb,255:red,92;green,92;blue,214}, from=1-1, to=1-3]
	\arrow["{\mu_J}", color={rgb,255:red,92;green,92;blue,214}, from=1-3, to=2-2]
	\arrow["{\exists! y}"', color={rgb,255:red,214;green,92;blue,214}, dotted, from=2-2, to=3-2]
	\arrow["{ \mu_{J'}}"', color={rgb,255:red,92;green,92;blue,214}, from=1-1, to=2-2]
	\arrow["{\nu_{J'}}"', color={rgb,255:red,214;green,92;blue,92}, curve={height=12pt}, from=1-1, to=3-2]
	\arrow["{\nu_J}", color={rgb,255:red,214;green,92;blue,92}, curve={height=-12pt}, from=1-3, to=3-2]
\end{tikzcd}\]
meaning that the \textcolor{rgb,255:red,92;green,92;blue,214}{blue} diagram commutes (for any \(J \subseteq J'\)) and that for any other commutative diagram as the \textcolor{rgb,255:red,214;green,92;blue,92}{red} one, there exists a unique morphism (in \textcolor{rgb,255:red,214;green,92;blue,214}{pink}) \(y: \Pi_F X \to Y\).

Then \(\cC\) is said to {\bf have filtered products} if any filter \(F\) and any $X = (X_i)_{i \in I}$ $I$-indexed family of objects in $\cC$ yield a filtered product of $X$ modulo \(F\).
\end{definition}

Filtered products being colimits of products, they are unique up to isomorphisms, allowing talking about \emph{the} filtered product.
To illustrate the definition, let us show that all presheaf categories $\widehat{\cB}$ have filtered products. The construction is a direct extension of the one in sets.

\begin{proposition}
\label{prop:filtered product in presheaves}
Let $\cB$ be a small category. The category of presheaves $\widehat{\cB}$ has filtered products.
\end{proposition}

\begin{proof}
Let $(G_i)_{i \in I}$ be a family of presheaves in $\widehat{\cB}$. Let $F$ be a filter on $I$. We construct the filtered product $\prod_F G$ of $(G_i)_{i \in I}$ modulo $F$ as: 
$$\prod_F G : \cB^{op} \to \Set; b \mapsto \prod_{i \in I} G_i(b)_{/_{\sim_F}}$$ 
where $\sim_F$ is the equivalence relation defined on the family of sets $(G_i(b))_{i \in I}$ by: $(a_i)_{i \in I} \sim_F (a'_i)_{i \in I} \Leftrightarrow \{i \in I \mid a_i = a'_i\} \in F$.

\medskip
For each $J \in F$, let $\mu_J : \prod_{j \in J} G_j \to \prod_F G$ defined for every $b \in |\cB|$ by the mapping $\mu_{J_b} : \prod_{j \in J} G_j(b) \to \prod_F G(b);a \mapsto [a']_{\sim_F}$ where $a = (a')_J$, i.e., \(a'\) is some extension of \(a\) to an \(I\)-indexed family. Because $F$ is a filter, $\mu_J$ is well-defined. Hence, the family $\mu = (\mu_J)_{J \in F}$ forms a cocone $A_F \Rightarrow \prod_F G$ where $A_F : F \to \widehat{\cB};J \mapsto \prod_{j \in J} G_j,J \subseteq J' \mapsto p_{J',J}$. Let $\nu : A_F \Rightarrow N$ be another cocone. For every $b \in |\cB|$, let us define the mapping $\theta_b : \prod_F G(b) \Rightarrow N(b)$ by: $[a]_{\sim_F} \mapsto \nu_{I_b}(a)$. 
It is not difficult to show that $\theta = (\theta_b)_{b \in |\cB|}$ is a natural transformation, and then $\mu : A_F \Rightarrow \prod_F \cM$ is indeed the colimit of $A_F : F \to \widehat{\cB}$.
\end{proof}

We now introduce a result on filtered products that we will use later.

\begin{proposition}
\label{prop: is an epimorphism}
Let $F$ be a filter on a set $I$ and $(X_i)_{i \in I}$ be a family of objects in $\cC$.
If \(p_{J',J} : \prod_{J'} \cM_j \to \prod_J \cM_j\) is an epimorphism for all inclusions \(J \subseteq J'\) in \(F\), then all coprojections \(\mu_J : \prod_J X_j \to \prod_F X\) (for \(J \in F\)) are epimorphisms.
\end{proposition}

\begin{proof}
Let \(F\) be a filter on a set \(I\).
For \(J \subseteq I\), \(F|_J = \set{J \cap K \mid K \in G}\) is a filter on \(J\), called the reduction of \(F\) to \(J\), satisfying \(\prod_F X \simeq \prod_{F|_J} X\) for any family \((X_i)_{i \in I}\) in \(\cC\) (see~\cite[Prop.~6.3]{Dia08}).

Now, let us fix a family \((X_i)_{i \in I}\) in \(\cC\) and \(J \in F\). Let \(f,g : \prod_F X \to Y\)
be two morphisms in \(\cC\) such that \(f \circ \mu_J = g \circ \mu_J\). By the previous isomorphism, we can consider \(f\) and \(g\) as morphisms \(\prod_{F|_J} X \to Y\). By hypothesis, for all \(K \subseteq J\), \(p_{J,K}\) is an epimorphism. In particular, for all \(K \in F|_J\), we obtain that \(f \circ \mu_K = g \circ \mu_K\). Since \(\prod_{F|_J} X\) is a colimit, the family \((\mu_K)_{K \in F|_J}\) is jointly epic. Therefore, \(f = g\) and \(\mu_J\) is an epimorphism.

\end{proof}

\subsubsection{Locally finitely presentable category}
\label{sec:lfp}

More abstractly, filtered products are an instance of colimit where the underlying diagram is a filtered category. We follow the presentation from~\cite{adamek_locally_1994}.

\begin{definition}[Filtered category]
    A {\bf filtered category} is a category \(\cC\) in which every finite diagram has a cocone.
\end{definition}

\begin{definition}[Filtered colimit]
    A {\bf filtered colimit} is a colimit of a functor \(\cD : \cI \to \cC\) where \(\cI\) is a filtered category.
\end{definition}

As discussed in~\cite{adamek_locally_1994}, directed (co)limits and filtered (co)limits are equivalent,\footnote{There is a mistake in proof of~\cite{adamek_locally_1994}, but a correct one appears in~\cite{andreka_direct_1982}, replacing finite subcategories with finite diagrams, and therefore unions with disjoint unions.} and a filtered product effectively corresponds to a filtered colimit and not filtered limit as the name would suggest. Filtered colimits enable the definition of finitely presentable objects.\footnote{Finitely presentable objects are called finitely presented in~\cite{Dia08} and also called compact in the literature.}

\begin{definition}[Finitely presentable object]
	\label{def:fp object}
    An object \(c\) of a category \(\cC\) is {\bf finitely presentable} if the hom-functor \(\Hom_\cC(c, \_) : \cC \to \Set\) preserves filtered colimits. For a category \(\cC\), we write \(\cC_{FP}\) for its full subcategory of finitely presentable objects. 
\end{definition}

This is equivalent to the following condition:
\begin{itemize}
    \item for every morphism $\mu : c \to d$ to the vertex of a colimiting co-cone $\nu : D \to d$ of a directed diagram $D : (I,\leq) \to \cC$, there exists $i \in I$ and a morphism $\mu_i : c \to D(i)$ such that $\mu = \nu_i \circ \mu_i$, and
    \item for any two morphisms $\mu_i$ and $\mu_j$ as above, there exists $k$ such that $k > i$, $k > j$, and $D_{i,k} \circ \mu_i = D_{j,k} \circ \mu_j$.
\end{itemize}

Many examples are given in~\cite{adamek_locally_1994}.

\begin{proposition}
    A finite colimit of finitely presentable objects is finitely presentable.
\end{proposition}

\begin{definition}[Locally finitely presentable category~\cite{adamek_locally_1994}]
	\label{def:lfp category}
    A locally small category\footnote{A category whose hom-sets are sets.} \(\cC\) is {\bf locally finitely presentable} if
    \begin{itemize}
        \item it has all small colimits (i.e., is cocomplete),
        \item it has a set \(A\) of finitely presentable objects such that every object in \(\cC\) is a filtered colimit of objects in \(A\).
    \end{itemize}
\end{definition}

Note that several equivalent definitions of locally finitely presentable categories might be found in the literature. For instance, the second condition can equivalently be stated as all isomorphism classes of objects in \(\cC_{FP}\) form a set (i.e., \(\cC_{FP}\) is skeletally small), and the restriction of the Yoneda embedding \(y : \cC \to \widehat{\cC_{FP}}\) defined by \(y(X) = \Hom_\cC(\_,X)\) is faithful and reflects isomorphisms (i.e., is conservative).

\section{Abstract categorical logic}
\label{sec:abstract categorical logic}

We now present the logical framework of~\cite{AB23} in which we will prove \Los's theorem (see Theorem~\ref{th:Los}). It essentially consists of 
\begin{itemize}
    \item a semantical system that enables abstract quantifiers to be defined as families of mappings induced by a context morphism, and
    \item a syntax based on an inductive construction of formulas over basic formulas directly interpretable by models.
\end{itemize}

\subsection{Semantical systems}
\label{subsec:semanticalsystems}

\begin{definition}[Semantical system]
\label{def:semantical system}
A {\bf semantical system} $\cS$ is given by:

\begin{itemize}
 	\item a prop-category $\cC$;
	\item a category $\Ctx$ whose objects are called {\bf contexts};
	\item a category of models $Mod$ with small products; 
	\item a functor $|\_| : Mod \to \cC^{\Ctx}$. %
\end{itemize}
\end{definition}

The categories \(\Ctx\) and \(Mod\) provide our categorical framework's expected generalizations of contexts and models. The objects of the category $\cC$ will denote model carriers and, therefore, interpret formulas through the functor $Prop_\cC$.
The functor \(|\_|\) relates these three categories, in such a way that \(|\cM|(\sigma)\) describes model carriers of \(\cM\) in \(\cC\), given the context \(\sigma\).
Compared to~\cite{AB23}, the category of models $Mod$ is now required to have small products such that we can consider filtered products.

\begin{proposition}
\label{prop:pullback functor natural}
For every context morphism $f : \sigma \to \tau$, the family
\[(|\cM|(f)^*:Prop_\cC(|\cM|(\tau)) \to Prop_\cC(|\cM|(\sigma)))_{\cM \in |Mod|}\]
is a natural transformation.   
\end{proposition}

\begin{proof}
Let $\mu: \cM \to \cM'$ be a model morphism and $f: \sigma \to \tau$ be a context morphism. As $|\mu|$ is a natural transformation, the following diagram commutes:
\[\begin{tikzcd}[column sep=4em]
	{|\cM|(\sigma)} & {|\cM|(\tau)} \\
	{|\cM'|(\sigma)} & {|\cM'|(\tau)}
	\arrow["{|\mu|_\tau}", from=1-2, to=2-2]
	\arrow["{|\mu|_\sigma}"', from=1-1, to=2-1]
	\arrow["{|\cM|(f)}", from=1-1, to=1-2]
	\arrow["{|\cM'|(f)}"', from=2-1, to=2-2]
\end{tikzcd}\]
As functors preserve commutative diagrams, we can apply \(Prop_\cC\) to the previous diagram, which yields that the following diagram
\[\begin{tikzcd}[column sep=4em]
	{Prop_\cC(|\cM|(\sigma))} & {Prop_\cC(|\cM|(\tau))} \\
	{Prop_\cC(|\cM'|(\sigma))} & {Prop_\cC(|\cM'|(\tau))}
	\arrow["{|\mu|_\tau^*}"', from=2-2, to=1-2]
	\arrow["{|\mu|_\sigma^*}", from=2-1, to=1-1]
	\arrow["{|\cM|(f)^*}"', from=1-2, to=1-1]
	\arrow["{|\cM'|(f)^*}", from=2-2, to=2-1]
\end{tikzcd}\]
is commutative.
\end{proof}

\subsubsection{Categorical first-order structures}
\label{fol-semanticalsystem}

Let $\Sigma = (S,F,R)$ be a multi-sorted first-order signature and \(V = (V_s)_{s \in S}\) an \(S\)-indexed family of variables. We define the semantical system $\cS_\Sigma = (\cC,\Ctx,Mod,|\_|)$ as follows:

\begin{itemize}

	\item $\cC$ is an elementary topos with small products provided with the functor \(\Sub : \cC^{op} \to \Heytalg\).\footnote{We could also have chosen any first-order hyperdoctrine of which a number of examples are given in~\cite{Pit02}.}

    \item $\Ctx$ is the category whose
		\begin{itemize}
			\item objects are all $\alpha$-equivalence classes $[\vec{x}]$ of finite sequences $\vec{x} = (x_1:s_1,\ldots,x_n:s_n)$ of distinct variables in \(V\), where $x_i:s_i$ means that the variable $x_i$ is of sort $s_i$;
   
			\item morphisms are the projections of $\alpha$-equivalence classes and the sequences of terms built using the signature \(\Sigma\). A morphism $\pi : [\vec{x}] \to [\vec{y}]$ is a projection if there exists $\vec{z}$ such that $\vec{x} = \vec{y}.\vec{z}$. A morphism $\vec{t} : [\vec{x}] \to [\vec{y}]$ with \(\vec{y} = (y_1:s_1,\ldots,y_n:s_n)\) and \(\vec{x} = (x_1:s'_1,\ldots,x_m:s'_m)\) is a sequence of first-order terms $(t_1:s_1,\ldots,t_n:s_n)$ built from the signature \(\Sigma\) if the variables of each term $t_i$ are in $\{x_1, \ldots, x_m\}$. 
		\end{itemize}

	\item $Mod$ is the category whose 
		\begin{itemize}
			\item objects are the $\Sigma$-structures $\cM$ defined by an $S$-indexed family of objects of $\cC$, i.e., for every $s \in S$, $M_s \in |\cC|$, and then for every $f:s_1 \times \ldots \times s_n \to s \in F$, $f^\cM:M_{s_1} \times \ldots \times M_{s_n} \to M_s$ is a morphism of $\cC$ and for every $r : s_1\times \ldots \times s_n \in R$, $r^\cM$ is a subobject in $\Sub(M_{s_1} \times \ldots \times M_{s_n})$, and
			\item morphisms between two $\Sigma$-structures $\cM$ and $\cN$ are families of morphisms $\mu = (\mu_s : M_s \to N_s)_{s \in S}$ such that:
			\begin{itemize}
				\item for every $f : s_1 \times \ldots \times s_n \to s \in F$, the diagram %
                \[\begin{tikzcd}
                	{M_{s_1} \times \ldots \times M_{s_n}} & {M_s} \\
                	{N_{s_1} \times \ldots \times N_{s_n}} & {N_s}
                	\arrow["{f^\cM}", from=1-1, to=1-2]
                	\arrow["{\mu_s}", from=1-2, to=2-2]
                	\arrow["{\mu_{s_1} \times \ldots \times \mu_{s_n}}"', from=1-1, to=2-1]
                	\arrow["{f^\cN}"', from=2-1, to=2-2]
                \end{tikzcd}\]
				commutes,
				\item for every $r : s_1 \times \ldots \times s_n \in R$, there exists a morphism \(O \to O'\) such that the diagram

                \[\begin{tikzcd}
                	O & {M_{s_1} \times \ldots \times M_{s_n}} \\
                	{O'} & {N_{s_1} \times \ldots \times N_{s_n}}
                	\arrow["{\mu_{s_1} \times \ldots \times \mu_{s_n}}", from=1-2, to=2-2]
                	\arrow["{r^\cM}", tail, from=1-1, to=1-2]
                	\arrow["{r^\cN}"', tail, from=2-1, to=2-2]
                	\arrow[from=1-1, to=2-1]
                \end{tikzcd}\]
                commutes.
			\end{itemize}
		\end{itemize} 
    
    Let us show that $Mod$ has small products. Let $(\cM_i)_{i \in I}$ be a family of models. Let $\prod_I \cM_i$ be the model defined by:
    \begin{itemize}
        \item for every $s \in S$, $(\prod_I M_i)_s = \prod_I M_{i_s}$,
        \item for every $f : s_1 \times \ldots \times s_n \rightarrow s \in F$, by the universal property of small products in \(\cC\),
        $f^{\prod_I \cM_i}$ is the unique morphism such that the following diagram
        \[\begin{tikzcd}[column sep=4em]
            {\prod_I M_{i_{s_1}} \times \ldots \times \prod_I M_{i_{s_n}}} & {\prod_I M_{i_s}} \\
            {M_{i_{s_1}} \times \ldots \times M_{i_{s_n}}} & {M_{i_s}}
            \arrow["{f^{\prod_I \cM_i}}", from=1-1, to=1-2]
            \arrow["{p_{I,i_{s_n}}}", from=1-2, to=2-2]
            \arrow["{p_{I,i_{s_1}} \times \ldots \times p_{I,i_{s_n}}}"', from=1-1, to=2-1]
            \arrow["{f^\cM_i}"', from=2-1, to=2-2]
        \end{tikzcd}\]	
    commutes for all \(i \in I\),
    \item for every $r:s_1 \times \ldots \times s_n \in R$, $r^{\prod_I \cM_i}$ is the subobject $\prod_I O_i \mto \prod_I M_{i_{s_1}} \times \ldots \times \prod_I M_{i_{s_n}}$ where $r^{\cM_i} : O_i \mto M_{i_{s_1}} \times \ldots \times M_{i_{s_n}}$.%

    \end{itemize}
    Since each \((\prod_I M_i)_s\) for \(s \in S\) is obtained as a small product in \(\cC\), it follows that $\prod_I \cM_i$ is the small product of $(\cM_i)_{i \in I}$.
    \item $|\_| : Mod \to \cC^{\Ctx}$ is the functor which:
    \begin{itemize}
        \item given a model $\cM$, associates the functor $|\cM| : \Ctx \to \cC$ defined:
        \begin{itemize}
            \item for every context $[\vec{x}]$ with $\vec{x} = (x_1:s_1,\ldots,x_n:s_n)$ by
            \[|\cM|([\vec{x}]) = M_{s_1} \times \ldots \times M_{s_n}\]
            \item for every projection $\pi : [\vec{x}] \to [\vec{y}]$ by the projection morphism
            \[|\cM|(\pi) : |\cM|([\vec{x}]) \to |\cM|([\vec{y}])\] 
            \item for every $\vec{t} = (t_1:s_1,\ldots,t_n:s_n) : [\vec{x}] \to [\vec{y}]$ with $\vec{y} = (y_1:s_1,\ldots,y_n:s_n)$ by the morphism
            \[(\sem{\cM}_{[\vec{x}]}(t_1),\ldots,\sem{\cM}_{[\vec{x}]}(t_n)) : |\cM|([\vec{x}]) \to |\cM|([\vec{y}])\] 
            where $\sem{\cM}_{[\vec{x}]}(t_i) : |\cM|([\vec{x}]) \to M_{s_i}$ is the interpretation of the term $t_i$ in the model $\cM$. Here, the interpretation of $t_i$ is defined inductively as follows:
            \begin{itemize}
                \item if \(t_i\) is a variable, it is an \(x_j\) in \(\vec{x}\), and \(\sem{\cM}_{[\vec{x}]}(t_i)\) is the projection on the sort associated with \(x_j\);
                \item if \(t_i\) is \(f(t'_1, \ldots t'_m)\) for some function symbol \(f\), with \(t'_1 : s'_1, \ldots t'_m:s'_m\) then \(\sem{\cM}_{[\vec{x}]}(t_i)\) is the composition
                \[|\cM|([\vec{x}]) \xrightarrow{(\sem{\cM}_{[\vec{x}]}(t'_1),\ldots,\sem{\cM}_{[\vec{x}]}(t'_m))} M_{s'_1} \times \ldots \times M_{s'_m} \xrightarrow{f^\cM} M_{s_i}\]
            \end{itemize}
        \end{itemize}
         
        \item given a morphism $\mu : \cM \to \cM' \in Mod$ associates the natural transformation $|\mu| : |\cM| \Rightarrow |\cM'|$ defined for every context $[\vec{x}]$ with $\vec{x} = (x_1:s_1,\ldots,x_n:s_n)$ by $|\mu|_{[\vec{x}]} = (\mu_{s_1},\ldots,\mu_{s_n})$. It is straightforward to show that for every $\gamma : [\vec{x}] \to [\vec{y}]$ the diagram
        \[\begin{tikzcd}[column sep=3em]
        	{|\cM|([\vec{x}])} & {|\cM'|([\vec{x}])} \\
        	{|\cM|([\vec{y}])} & {|\cM'|([\vec{y}])}
        	\arrow["{|\mu|_{[\vec{x}]}}", from=1-1, to=1-2]
        	\arrow["{|\cM'|(\gamma)}", from=1-2, to=2-2]
        	\arrow["{|\cM|(\gamma)}"', from=1-1, to=2-1]
        	\arrow["{|\mu|_{[\vec{y}]}}"', from=2-1, to=2-2]
        \end{tikzcd}\]
        commutes.
    \end{itemize} 
\end{itemize}
\subsubsection{Categorical higher-order structures}
\label{hol-semanticalsystem}

Given a multi-sorted first-order signature $\Sigma = (S,F,R)$, profiles of functions and relations are now types whose set, denoted $\Sigma$-Typ, is inductively defined as follows: 

\begin{itemize}
	\item {\em Basic types.} $S \subseteq \mbox{$\Sigma$-Typ}$;
	\item {\em Product types.} If $A,B \in \mbox{$\Sigma$-Typ}$, then $A \times B \in \mbox{$\Sigma$-Typ}$;
	\item {\em Function types.} If $A,B \in \mbox{$\Sigma$-Typ}$, then $B^A \in \mbox{$\Sigma$-Typ}$; 
	\item {\em Power types.} If $A \in \mbox{$\Sigma$-Typ}$, then $PA \in \mbox{$\Sigma$-Typ}$. 
\end{itemize} 
Now, functions in $F$ and relations in $R$ have profiles in $\Sigma$-Typ. Hence, a function $f \in F$ has a profile defined by an ordered pair $(A,B)$ of $\Sigma$-Typ, and we write $f : A \to B$ to mean that $f$ has the profile $(A,B)$. Likewise, a relation $r \in R$ has a profile defined by a type $A \in \mbox{$\Sigma$-Typ}$.
 
The semantical system $\cS_\Sigma$ for interpreting higher-order categorical logic is then any tuple $(\cC,\Ctx,Mod,|\_|)$ defined as in Section~\ref{fol-semanticalsystem} with the condition that we have a stock of variables $x:A$ for each type $A \in \mbox{$\Sigma$-Typ}$, and the fact that model carriers are extended to elements in $\Sigma$-Typ.

\subsubsection{First-order institutions}
\label{subsubsec:semantical system for institutions}

Although in institution theory we deal with closed formulas (also called sentences), open formulas can also be considered through signature morphisms. Indeed, a set of variables can be identified with a signature extension (variables are then treated as constants), and the valuation of variables in a model is just a model expansion along with signature morphisms.
We show that we can define a semantical system for each signature on which we define the usual first-order quantifiers. Before doing so, let us recall the definition of an institution.
\begin{definition}[Institution~\cite{GB92}]
An {\bf institution} $\mathcal{I} = (Sig,Sen,Mod,\models)$ consists of
\begin{itemize}
\item a category $Sig$ whose objects are called {\em signatures} and are denoted by $\Sigma$,
\item a functor $Sen : Sig \rightarrow \Set$ giving for each signature $\Sigma$ a set $Sen(\Sigma)$ whose elements are called {\em sentences}, 
\item a contravariant functor $Mod : Sig^{op} \rightarrow Cat$ giving for each signature its category of models, and 
\item a $Sig$-indexed family of relations $\models_\Sigma \subseteq |Mod(\Sigma)| \times Sen(\Sigma)$ called 
{\em satisfaction relation}, such that the following property, called the {\em satisfaction condition}, holds: for all $\theta : \Sigma \rightarrow \Sigma'$ in $Sig$, for all $\cM'$ in 
$|Mod(\Sigma')|$, and for all $\varphi$ in $Sen(\Sigma)$,
\[
    \mathcal{M}' \models_{\Sigma'} Sen(\theta)(\varphi) \Longleftrightarrow Mod(\theta)(\mathcal{M}') \models_\Sigma \varphi
\]
\end{itemize}
\end{definition}

We do not give examples here because many are described in~\cite{Dia08}, and we refer readers interested in such examples to this book.  
A supplementary condition is needed to capture the notion of FOL quantification: {\em quasi-representable signature morphism}~\cite[Chap. 5, p. 102]{Dia08}. A signature morphism $\chi : \Sigma \to \Sigma'$ is {\em quasi-representable} if for each $\Sigma'$-model $\cM'$,
the following isomorphism (of comma categories) holds
\[
    \cM'/Mod(\Sigma') \simeq Mod(\chi)(\cM')/Mod(\Sigma).
\]
Then, given a morphism $\mu : Mod(\chi)(\cM') \to \cN$ in \(Mod(\Sigma)\), the morphism $\mu' : \mathcal{M}' \to \mathcal{N}'$ induced by the above bijection is the {\em $\chi$-expansion} of \(\mu\).
Note that in FOL, each signature extension with constants is quasi-representable, but any morphism extending the signature with either relation or non-constant function symbol is not quasi-representable.

In the sequel, we request two additional properties for institutions (to handle filtered products) and only consider institutions $\mathcal{I} = (Sig,Sen,Mod,\models)$ such that:
\begin{itemize}
    \item for every signature $\Sigma \in |Sig|$, $Mod(\Sigma)$ is with small products.
    \item for every signature morphism $\chi$, the forgetful functor $Mod(\chi)$ creates small products.
\end{itemize}
We define the semantical system $\mathcal{S}_\mathcal{I}(\Sigma) = (\cC,\Ctx,Mod,|\_|)$ as follows:
\begin{itemize}
    \item $\cC$ is the category whose 
    \begin{itemize}
        \item objects are the sets \(|\mathcal{M}|(\chi)\) defined for every object $\mathcal{M}$ of \(Mod(\Sigma)\) and every quasi-representable signature morphism $\chi : \Sigma \to \Sigma'$ as
        \[
            |\mathcal{M}|(\chi) = \{\mathcal{M}' \in |Mod(\Sigma')| \mid Mod(\chi)(\mathcal{M}') = \mathcal{M}\}
        \] 
        \item morphisms are the mappings 
        \[(\mu,\theta) : 
        \left\{
        \begin{array}{ccc}
        |\mathcal{M}|(\chi_2) & \to & |\cN|(\chi_1) \\
        \mathcal{M}_2 & \mapsto & Mod(\theta)(\codom(\mu')) 
        \end{array}
        \right.\]
        where $\theta : \Sigma_1 \to \Sigma_2$ is a signature morphism such that $\theta \circ \chi_1 = \chi_2$ with $\chi_i :\Sigma \to \Sigma_i$ (for $i \in \{1,2\}$), and $\mu'$ is the unique $\chi_2$-expansion of $\mu : \mathcal{M} \to \cN$.
    \end{itemize}

    Let us detail the construction of \((\mu,\theta)\). We consider a morphism \(\mu: \cM \to \cN\) in \(Mod(\Sigma)\) and a morphism \(\theta: \Sigma_1 \to \Sigma_2\) in \(Sig\) such that the following triangle commutes.
    \[\begin{tikzcd}[row sep=tiny]
    	& {\Sigma_1} \\
    	\Sigma \\
    	& {  \Sigma_2}
    	\arrow["{\chi_1}", from=2-1, to=1-2]
    	\arrow["\theta", from=1-2, to=3-2]
    	\arrow["{\chi_2}"', from=2-1, to=3-2]
    \end{tikzcd}\]
    Applying the functor \(Mod\) yields the following commutative triangle.
    \[\begin{tikzcd}[row sep=tiny]
    	& {Mod(\Sigma_1)} \\
    	{Mod(\Sigma)} \\
    	& {  Mod(\Sigma_2)}
    	\arrow["{Mod(\chi_1)}"', from=1-2, to=2-1]
    	\arrow["{Mod(\theta)}"', from=3-2, to=1-2]
    	\arrow["{Mod(\chi_2)}", from=3-2, to=2-1]
    \end{tikzcd}\]
    Then \(\cM_2\) in \(|\cM|(\chi_2)\) is a model in \(|Mod(\Sigma_2)|\) such that \(Mod(\chi_2)(\cM_2) = \cM\). We can therefore reinterpret \(\mu\) as a morphism \(Mod(\chi_2)(\cM_2) \to \cN\) and consider its unique \(\chi_2\)-expansion \(\mu': \cM_2 \to \cN'\) where \(\cN = \codom(\mu)\) is a model in \(|Mod(\Sigma_2)|\). Applying \(Mod(\theta)\) to \(\cN\) yields an element of \(|\cN(\chi_1)|\) which we consider as the image of \(\cM_2\) by \((\mu,\theta)\).

    \medskip
    Since objects in \(\cC\) are sets and morphisms are functions, \(Prop_\cC\) is the contravariant power set functor.
    In other words, given an object $|\mathcal{M}|(\chi) \in |\cC|$, $Prop(|\mathcal{M}|(\chi)) = \powerset(|\mathcal{M}|(\chi))$, and given a morphism $(\mu,\theta)$ in $\cC$, $(\mu,\theta)^*$ is the mapping $S' \mapsto (\mu,\theta)^{-1}(S')$, i.e.
    \[(\mu,\theta)^* : S' \mapsto 
    \left\{
    \begin{array}{l|l}
    & Mod(\chi_2)(\mathcal{M}_2) = \mathcal{M}, \\
    \mathcal{M}_2  & \mbox{and } Mod(\theta)(\codom(\mu')) \in S', \\ 
    & \mbox{where} \; \mu' \; \mbox{unique $\chi_2$-expansion of} \; \mu
    \end{array}
    \right\}
    \]    

    \item \(Prop_\cC\) is the contravariant powerset functor \(\powerset\) from Section~\ref{sec:elementary topos} (as we are dealing with sets).

    \item $\Ctx$ is the opposite of the full subcategory of the under (or co-slice) category $\Sigma / Sig$ where objects are all quasi-representable morphisms;
    \item $Mod = Mod(\Sigma)$; and 
    \item $|\_| : Mod \to \cC^{\Ctx}$ is the functor which
    \begin{itemize}
        \item to every $\mathcal{M} \in |Mod|$ associates the functor
        \[|\mathcal{M}| : \Ctx \to \cC;\chi \mapsto |\mathcal{M}|(\chi);\theta \mapsto (Id_\mathcal{M},\theta)\]
        \item to every $\mu : \mathcal{M} \to \cN$ associates the natural transformation
        \[|\mu| : |\mathcal{M}| \Rightarrow |\cN|\]
        defined for every quasi-representable signature morphism $\chi : \Sigma \to \Sigma'$ by the mapping $|\mu|_\chi = (\mu,Id_{\Sigma'})$. Let $\theta : \chi_2 \to \chi_1$ be a morphism in $\Ctx$. It is quite straightforward to show that the diagram
        \[\begin{tikzcd}[column sep=3em]
        	{|\cM|(\chi_2)} & {|\cM'|(\chi_2)} \\
        	{|\cM|(\chi_1)} & {|\cM'|(\chi_1)}
        	\arrow["{|\cM|(\theta)}"', from=1-1, to=2-1]
        	\arrow["{|\mu|_{\chi_1}}"', from=2-1, to=2-2]
        	\arrow["{|\mu|_{\chi_2}}", from=1-1, to=1-2]
        	\arrow["{|\cM'|(\theta)}", from=1-2, to=2-2]
        \end{tikzcd}\]
        commutes.
    \end{itemize}
\end{itemize}

\subsubsection[Modal logic (ML) and co-algebras]{Modal logic (ML) and co-algebras~\cite{Jacobs16,Rut00}}
\label{modal-semanticalsystem}

Let $\cC$ be an elementary topos with small products. 
Let $F : \cC \to \cC$ be a functor which preserves small products.\footnote{In practice, and most of the time in computer science, the category $\cC$ is $\Set$~\cite{Rut00}.} Hence, the semantical system $\cS_F = (\cC,\Ctx,Mod,|\_|)$ is defined by:

\begin{itemize}
	\item $\Ctx$ is the trivial category with a unique object written $\bullet$; 
	\item $Mod$ is the category whose objects are $F$-coalgebras defined by pairs $(X,\alpha_X : X \to F(X))$ where $X \in |\cC|$ and $\alpha_X$ is a morphism of $\cC$, and a morphism between two $F$-coalgebras $(X,\alpha_X)$ and $(Y,\alpha_Y)$ is a morphism $\mu : X \to Y$ such that the following diagram 
    \[\begin{tikzcd}
    	X & Y \\
    	{F(X)} & {F(Y)}
    	\arrow["\mu", from=1-1, to=1-2]
    	\arrow["{\alpha_Y}", from=1-2, to=2-2]
    	\arrow["{\alpha_X}"', from=1-1, to=2-1]
    	\arrow["{F(\mu)}"', from=2-1, to=2-2]
    \end{tikzcd}\]
commutes.

\medskip
Let us show that $Mod$ has small products.\footnote{This is an instantiation of a more general result which means that any type of limit that is preserved by $F$ also exists in the category of $F$-coalgebras~\cite{Rut00}.}
Let $((X_i, \alpha_{X_i}))_{i \in I}$ be a family of $F$-coalgebras. Since \(\cC\) has small products, \(\prod_I X_i\) uniquely exists in \(\cC\) (with the associated projections), which directly extends to \(\prod_I F(X_i)\) because $F$ preserves small products. We can then define \(\alpha_{\prod_I X_i}\) by the universal property of small products, i.e., as the unique morphism such that
\[\begin{tikzcd}
    	\prod_I X_i & X_i \\
    	{\prod_I F(X_i)} & {F(X_i)}
    	\arrow["p_{I,i}", from=1-1, to=1-2]
    	\arrow["{\alpha_{X_i}}", from=1-2, to=2-2]
    	\arrow["{\alpha_{\prod_I X_i}}"', from=1-1, to=2-1]
    	\arrow["{F(p_{I,i})}"', from=2-1, to=2-2]
    \end{tikzcd}\]
commutes for each \(i \in I\). Then \(\prod_I(X_i,\alpha_{X_i}) = (\prod_I X_i, \alpha_{\prod_I X_i})\).

	\item $|\_| : Mod \to \cC^{\Ctx}$ is the functor which to every coalgebra $(X,\alpha_X)$ associates the functor $|(X,\alpha_X)| : \bullet \mapsto X$, and to every morphism $\mu : (X,\alpha_X) \to (Y,\alpha_Y)$ associates the natural transformation $|\mu| : |(X,\alpha_X)| \Rightarrow |(Y,\alpha_Y)|$ defined by $|\mu|_\bullet = \mu$.
\end{itemize}

\subsection{Quantifiers}

\begin{definition}[Quantifier]
\label{def:quantifier}
Let $\cS = (\cC,\Ctx,Mod,|\_|)$ be a semantical system. Let $f : \sigma \to \tau$ be a morphism in $\Ctx$. For $n \in \mathbb{N}$, a {\bf n-ary quantifier over $f$} is a family $(\cQ f_\cM)_{\cM \in |Mod|}$ where for every $\cM \in |Mod|$, $\cQ f_\cM : Prop_\cC^n(|\cM|(\sigma)) \to Prop_\cC(|\cM|(\tau))$ is a mapping,\footnote{$Prop_\cC^n : \cC^{op} \to Pos;X \mapsto Prop_\cC(X) \times \ldots \times Prop_\cC(X)$ is the $n$-fold product of the contravariant functor $Prop_\cC$.} isotone in every argument for the orders $\preceq_{|\cM|(\sigma)}$ and $\preceq_{|\cM|(\tau)}$.
\end{definition}

\begin{definition}[Distributing quantifier]
\label{def:distributing over}
A quantifier $\cQ f$ is {\bf distributing} over a model morphism $\mu : \cM \to \cM'$ if the following diagram
\[\begin{tikzcd}
	{Prop^n_\cC(|\cM|(\sigma))} & {Prop_\cC(|\cM|(\tau))} \\
	{Prop^n_\cC(|\cM'|(\sigma))} & {Prop_\cC(|\cM'|(\tau))}
	\arrow["{\cQ f_\cM}", from=1-1, to=1-2]
	\arrow["{|\mu|^*_\tau}"', from=2-2, to=1-2]
	\arrow["{(|\mu|^*_\sigma,\ldots,|\mu|^*_\sigma)}", from=2-1, to=1-1]
	\arrow["{\cQ f_{\cM'}}"', from=2-1, to=2-2]
\end{tikzcd}\]
commutes.
\end{definition}

We could have required quantifiers to be natural transformations, but this is rarely verified except for certain types of model morphisms (essentially certain families of epimorphisms). This distributivity property will be useful to prove \Los's result.

\subsubsection{First-order and higher-order quantifiers}
\label{fol-quantifiers}
Elementary toposes interpret first-order and higher-order logics because the functor $\Sub : \cC^{op} \to \Heytalg$ is a tripos over $\cC$. 
In other words, for every morphism $f : X \to Y$ in an elementary topos $\cC$, the pullback functor $f^* : \Sub(Y) \to Sub(X)$ has a right-adjoint $\forall f : \Sub(X) \to \Sub(Y)$ and a left-adjoint $\exists f : \Sub(X) \to \Sub(Y)$.
So, given a morphism $\vec{t} : [\vec{x}] \to [\vec{y}]$, $\forall \vec{t}$ and $\exists \vec{t}$ are
defined for every model by $\cM$ by: $\forall \vec{t}_\cM = \forall \sem{\cM}_{[\vec{x}]}(\vec{t})$ and $\exists \vec{t}_\cM = \exists \sem{\cM}_{[\vec{x}]}(\vec{t})$, where \(\sem{\cM}_{[\vec{x}]}(\vec{t})\) is the interpretation defined in Section~\ref{fol-semanticalsystem}.
More details on these quantifiers can be found in~\cite{AB23}.

\medskip
In later proofs, we will use the exact expressions of the quantifiers for the projection morphisms, which we now give. Let $\pi : [\vec{x}] \to [\vec{y}]$ be a projection morphism in $\Ctx$ (with \(\vec{x} = \vec{y}.\vec{z}\) for some finite sequence of variables \(\vec{z}\)). Using the internal language of $\cC$ (see Appendix~\ref{sec:AppB}), the morphisms $\forall \pi_\cM, \exists \pi_\cM : P|\cM|([\vec{x}]) \to P|\cM|([\vec{y}])$ yield
\[\forall \pi_{\cM}(S) = \{y : |\cM|([\vec{y}]) \mid \forall z:|\cM|([\vec{z}]), (y,z) \in_{|\cM|([\vec{x}])} S\}\]
\[\exists \pi_{\cM}(S) = \{y : |\cM|([\vec{y}]) \mid \exists z:|\cM|([\vec{z}]), (y,z) \in_{|\cM|([\vec{x}])} S\}\]
where $S:P |\cM|([\vec{x}])$ is a variable. 

\begin{proposition}
\label{prop:distributing in FOL}
Let \(\mu: \cM \to \cM'\) be a model morphism such that for every context \([\vec{x}]\) in \(\Ctx\), \(|\mu|_{[\vec{x}]}\) is an epimorphism,\footnotemark{} then FOL quantifiers are distributing over \(\mu\).

\footnotetext{%
Let us recall that epimorphisms in toposes are regular. A regular epimorphism is a morphism $\mu : X \to Y$ (in a given category) that is the coequalizer of some parallel pair of morphisms. The usefulness of such a morphism is that in a regular category, it satisfies:
\[
    \forall y:Y,\exists x:X,\mu(x) = y
\]
}

\end{proposition}

\begin{proof}
Let \(\mu: \cM \to \cM'\) be a model morphism such that for every context \([\vec{x}]\) in \(\Ctx\), \(|\mu|_{[\vec{x}]}\) is an epimorphism.
Let $\pi : [\vec{x}] \to [\vec{y}]$ be a projection morphism in $\Ctx$ (with \(\vec{x} = \vec{y}.\vec{z}\) for some \(\vec{z}\)).
Finally, let \(X' : P |\cM'|([\vec{x}])\) be a variable.

We detail the case of universal quantifiers.
From the expressions of \(\forall \pi_\cM\) and \(|\mu|^*_{[\vec{x}]}\) in the internal logic of elementary toposes (where \(\cM\) is a model and \([\vec{x}]\) a context), we obtain
\[
    \forall \pi_{\cM} \circ |\mu|^*_{[\vec{x}]}(X') = \{ m_y : |\cM|([\vec{y}]) \mid \forall m_z : |\cM|([\vec{z}]),~ |\mu|_{[\vec{x}]}(m_y,m_z) \in_{|\cM'|([\vec{x}])} X' \}
\]
and 
\[
    |\mu|^*_{[\vec{y}]} \circ \forall \pi_{\cM'}(X') = \{ m_y : |\cM|([\vec{y}]) \mid \forall m'_z : |\cM'|([\vec{z}]),~
    (|\mu|_{[\vec{y}]}(m_y),m'_z) \in_{|\cM'|([\vec{x}])} X' \}
\]

Let \(m_y : |\cM|([\vec{y}])\) be a variable such that \(m_y \in_{|\cM|([\vec{y}])} |\mu|^*_{[\vec{y}]} \circ \forall \pi_{\cM'}(X')\).
Let \(m_z : |\cM|([\vec{z}])\) be a variable.
Then \(|\mu|_{[\vec{x}]}(m_y,m_z) = (|\mu|_{[\vec{y}]}(m_y),|\mu|_{[\vec{z}]}(m_z))\).
Since \(|\mu|_{[\vec{z}]}(m_z)\) is some \(m'_z : |\cM'|([\vec{z}])\), it follows that \(|\mu|_{[\vec{x}]}(m_y,m_z) \in_{|\cM'|([\vec{x}])} X'\), i.e., that \(m_y \in_{|\cM|([\vec{y}])} \forall \pi_{\cM} \circ |\mu|^*_{[\vec{x}]}(X')\).

Conversely, let \(m_y : |\cM|([\vec{y}])\) be a variable such that \(m_y \in_{|\cM|([\vec{y}])} \forall \pi_{\cM} \circ |\mu|^*_{[\vec{x}]}(X')\)
Let \(m'_z : |\cM'|([\vec{z}])\) be a variable.
Since \(|\mu|_{[\vec{z}]}\) is an epimorphism, there exists \(m_z : |\cM|([\vec{z}])\) such that \(|\mu|_{[\vec{z}]}(m_z) = m'_z\).
Then, \(|\mu|_{[\vec{x}]}(m_y,m_z) \in_{|\cM'|([\vec{x}])} X'\), which implies that \((|\mu|_{[\vec{y}]}(m_y),m'_z) \in_{|\cM'|([\vec{x}])} X'\), i.e., that \(m_y \in_{|\cM|([\vec{y}])} |\mu|^*_{[\vec{y}]} \circ \forall \pi_{\cM'}(X')\).

\medskip
For existential quantifiers, starting from \(m_z : |\cM|([\vec{y}])\), \(|\mu|_{[\vec{z}]}(m_z)\) provides a variable of \(|\cM'|([\vec{y}])\) meaning that any \(m_y \in_{|\cM|([\vec{y}])} \exists \pi_{\cM} \circ |\mu|^*_{[\vec{x}]}(X')\) satisfies \(m_y \in_{|\cM|([\vec{y}])} |\mu|^*_{[\vec{y}]} \circ \exists \pi_{\cM'}(X')\). Similarly to the universal quantifier, the fact that \(|\mu|_{[\vec{z}]}\) is an epimorphisms allows retrieving an \(m_z : |\cM|([\vec{y}])\) from an \(m'_z : |\cM'|([\vec{y}])\) such that \(|\mu|_{[\vec{z}]}(m_z) = m'_z\), which yields that any \(m_y \in_{|\cM|([\vec{y}])} |\mu|^*_{[\vec{y}]} \circ \exists \pi_{\cM'}(X')\) satisfies \(m_y \in_{|\cM|([\vec{y}])} \exists \pi_{\cM} \circ |\mu|^*_{[\vec{x}]}(X')\).

\medskip
The proof extends to quantifiers over any context morphism.
\end{proof}

\medskip
These constructions directly extend to HOL quantifiers since the category of contexts is essentially the same.

By Proposition~\ref{prop: is an epimorphism} and Proposition~\ref{prop:distributing in FOL}, under the hypothesis that for every $J \subseteq J' \in F$, $p_{J,J'}$ is an epimorphism, we have that FOL quantifiers are distributing over $p_{J',J}$ and $\mu_J$.

\subsubsection{First-order quantifiers in institutions}
\label{institutions-quantifiers}
Institutions admit FOL quantifiers for any morphism $\theta : \chi_1 \to \chi_2$ in $\Ctx$ for the semantical system given in Section~\ref{subsubsec:semantical system for institutions}. We recall their expressions next, while additional details may be consulted in~\cite{AB23}.
Given a morphism $\theta : \chi_1 \to \chi_2$ and a model $\cM \in |Mod(\Sigma)|$, $\forall \theta_{\cM}$ and $\exists \theta_{\cM}$ are defined for every $S \subseteq |\cM|(\chi_2)$ as follows:
\begin{align*}
    \forall \theta_{\cM}(S) &= \{\cM_1 \in |\cM|(\chi_1) \mid \forall \cM_2 \in |Mod(\Sigma_2)|, Mod(\theta)(\cM_2) = \cM_1 \text{ implies } \cM_2 \in S \}\\
    \exists \theta_{\cM}(S) &= \{\cM_1 \in |\cM|(\chi_1) \mid \exists \cM_2 \in |Mod(\Sigma_2)|, Mod(\theta)(\cM_2) = \cM_1 \text{ and } \cM_2 \in S \}
\end{align*}

\begin{proposition}
\label{prop:distribution in institutions}
Institution FOL quantifiers are distributing over any morphism $\mu : \cM \to \cN$ which satisfies the following property: 

\begin{itemize}
    \item {\bf Existence of extension.} for every quasi-representable morphism $\chi : \Sigma \to \Sigma'$, and every model $\cN' \in |Mod(\Sigma')|$ such that $Mod(\chi)(\cN') = \cN$, there exists a morphism  $\mu' : \cM' \to \cN'$ which is the unique $\chi$-expansion of $\mu$ (and then $Mod(\chi)(\cM') = \cM$).
\end{itemize}

\end{proposition}

In $FOL$ over presheaves, this holds when morphisms are epic (i.e., for every $s \in S$, for every $b \in |\cB|$, $\mu_{s_b} : M_s(b) \to M_s(b)$ is a surjective mapping). 

\medskip
\begin{proof}
For every such morphism \(\mu : \cM \to \cN\) and every $\theta : \chi_2 \to \chi_1 \in |\Ctx|$, the following diagram 
\[\begin{tikzcd}
	{\powerset(|\cM|(\chi_2))} & {\powerset(|\cM|(\chi_1))} \\
	{\powerset(|\cN|(\chi_2))} & {\powerset(|\cN|(\chi_1))}
	\arrow["{\forall \theta_\cM}", from=1-1, to=1-2]
	\arrow["|\mu|^*_{\chi_1}"', from=2-2, to=1-2]
	\arrow["|\mu|^*_{\chi_2}", from=2-1, to=1-1]
	\arrow["{\forall \theta_{\cN}}"', from=2-1, to=2-2]
\end{tikzcd}\]
commutes. Indeed, unfolding the definitions of \(\forall \theta\) and \(|\mu|^*\), we obtain the following formulas for \(|\mu|^*_{\chi_1}(\forall \theta_{\cN}(S))\) and \(\forall \theta_\cM(|\mu|^*_{\chi_2}(S))\), where \(S\) is a subset of \(|\cN|(\chi_2)\):
\[
    |\mu|^*_{\chi_1}(\forall \theta_{\cN}(S)) = \{\cM_1 \in |\cM|(\chi_1) \mid \forall \cN_2 \in |Mod(\Sigma_2)|,~
    Mod(\theta)(\cN_2) = \codom(\mu_1) \mbox{ implies } \cN_2 \in S\}
\]
where $\mu_1$ is the unique $\chi_1$-expansion of $\mu : Mod(\chi_1)(\cM_1) \to \cN$, and
\[
    \forall \theta_\cM(|\mu|^*_{\chi_2}(S)) = \{\cM_1 \in |\cM|(\chi_1) \mid \forall \cM_2 \in |Mod(\Sigma_2)|,~
    Mod(\theta)(\cM_2) = \cM_1 \mbox{ implies } \codom(\mu_2) \in S\}
\]
where $\mu_2$ is the unique $\chi_2$-expansion of $\mu : Mod(\chi_1)(Mod(\theta)(\cM_2)) \to \cN$.

\medskip
Let $\cM_1 \in |\mu|^*_{\chi_1}(\forall \theta_{\cN}(S))$. Let $\cM_2 \in Mod(\Sigma_2)$ such that $Mod(\theta)(\cM_2) = \cM_1$. Then it follows that $\mu : Mod(\chi_1)(Mod(\theta)(\cM_2)) \to \cN$ and we can consider $\mu_1 : Mod(\theta)(\cM_2) \to \codom(\mu_1)$, the unique $\chi_1$-expansion of $\mu$. Now, let $\mu_2 : \cM_2 \to \codom(\mu_2)$ be the unique $\chi_2$-expansion of $\mu$. By the uniqueness of extensions, we have that $\mu_2$ is the unique $\theta$-expansion of $\mu_1$, and then $Mod(\theta)(\codom(\mu_2)) = \codom(\mu_1)$ from which we can conclude that $\codom(\mu_2) \in S$.

\medskip
Let $\cM_1 \in \forall \theta_\cM(|\mu|^*_{\chi_2}(S))$. Let $\cN_2 \in Mod(\Sigma_2)$ such that $Mod(\theta)(\cN_2) = \codom(\mu_1)$ where $\mu_1 : \cM_1 \to \codom(\mu_1)$ is the unique $\chi_1$-expansion of $\mu : Mod(\chi_1)(\cM_1) \to \cN$. By the property of existence of extension, there exists a morphism $\mu_2 : \cM_2 \to \cN_2$, which is the unique $\theta$-expansion of $\mu_1$, and then of $\mu$. Hence, we have that $Mod(\theta)(\cM_2) = \cM_1$, from which we can conclude that $\cN_2 \in S$. 

\medskip
From the duality of $\exists \theta$ with $\forall \theta$ (i.e., $\forall \theta_\cM(S) = \exists \theta_\cM(S^c)^c$), we directly have that
\[
    \exists \theta_\cM \circ |\mu|^*_{\chi_2} = |\mu|^*_{\chi_1} \circ \exists \theta_{\cN}
\]
\end{proof}

Once again, FOL existential quantifiers are weakly distributing over any model morphism.

\subsubsection{Modalities in coalgebraic logic}
\label{modal:quantifiers}
In the framework of coalgebras, the notion of predicate lifting has been identified as the concept underlying modal operator semantics~\cite{Pat03}. Let us place ourselves in the semantical system of Section~\ref{modal-semanticalsystem}. Let $F : \cC \to \cC$ be a functor. A {\em $n$-ary predicate lifting} is then a natural transformation $\lambda : \cP^n \Rightarrow \cP \circ F^{op}$ such that
for every $X \in |\cC|$, $\lambda_X$ is preserving orders.
In other words, for all variables $x_1,\dots,x_n,x'_1,\ldots,x'_n:X$ such that \(x_1 \preceq_X x'_1\), \ldots, \(x_n \preceq_X x'_n\), then \(\lambda_X(x_1,\ldots,x_n) \preceq_{F(X)} \lambda_X(x'_1,\ldots,x'_n)\).

Given a \(F\)-coalgebra \((X,\alpha_X)\), one can (internally) define a morphism \(\alpha^{-1}_X: PF(X) \to PX\) such that for a variable \(Y:PF(X)\),
\[
    \alpha^{-1}_X(Y) = \{x:X \mid \alpha_X(x) \in_{F(X)} Y\}
\]

In the category $\Ctx$, the only morphism is $Id_{\bullet}$, then given an $n$-ary predicate lifting $\lambda$ and a co-algebra $(X,\alpha_X)$, we can internally define $[\lambda]_X : P^nX \to PX$ by: $[\lambda]_X = \alpha^{-1}_X \circ \lambda_X$. 

\begin{proposition}
The family $[\lambda] = ([\lambda]_X)_{X \in |\cC|}$ is a natural transformation. 
\end{proposition}

\begin{proof}
Since \(\lambda\) is a natural transformation, it suffices to show that $\alpha^{-1} : \cP \circ F \Rightarrow \cP$ is a natural transformation. Let $\mu : X \to X'$ be a morphism of coalgebras. From the definition of $[\lambda]_X$, showing that $\alpha^{-1}$ is a natural transformation amounts to show that: 
$$\cP(\mu) \circ \alpha^{-1}_{X'} = \alpha^{-1}_X \circ \cP(F(\mu))$$
Let $Y':PF(X')$ be a variable. By definition, we have the two following equations:
\begin{align*}
\cP(\mu)(\alpha^{-1}_{X'}(Y')) &= \{x:X \mid \alpha_{X'}(\mu(x)) \in_{F(X')} Y'\} \\
\alpha^{-1}_X(\cP(F(\mu))(Y')) &= \{x:X \mid F(\mu)(\alpha_X(x)) \in_{F(X')} Y'\}
\end{align*}
Since $\mu$ is a coalgebra morphism \(F(\mu)(\alpha_X(x)) = \alpha_{X'}(\mu(x))\) for any \(x : X\), which yields the desired commutative property.
\end{proof}

\begin{corollary}
    The family $[\lambda]$ is distributing over any model morphism.
\end{corollary}

\begin{corollary}
    If $\lambda$ is an $n$-ary predicate lifting, so is $[\lambda]$.
\end{corollary}

\begin{proof}
    It is quite simple to show that $\alpha^{-1}_X$ is isotone, and then $[\lambda]_X$ is isotone on its arguments since $\lambda_X$ is.
\end{proof}

\medskip
We now present two examples of isotone 1-ary predicate liftings~\cite{Pat03}. Let us suppose that we have a natural transformation $\beta : F \Rightarrow \exists$, where \(\exists : \cC \to \cC\) is the covariant functor introduced in Section~\ref{sec:elementary topos}. Then, we internally define the two following predicate liftings $\forall \beta,\exists \beta : \exists \Rightarrow \exists \circ F$ as

\[\forall \beta_X(Y) = \{y:F(X) \mid \beta_X(y) \preceq_X Y\}\]
\[\exists \beta_X(Y) = \{y:F(X) \mid \exists x:X, x \in_X \beta_X(y) \wedge x \in_X Y\}\]
where $Y:PX$ is a variable.

Showing that $\forall \beta = (\forall \beta_X)_{X \in |\cC|}$ and $\exists \beta = (\exists \beta_X)_{X \in |\cC|}$ are isotone for every $X \in |\cC|$ is quite straightforward. Then, $\forall_X = \alpha^{-1}_X \circ \forall \beta_X$ and $\exists_X = \alpha^{-1}_X \circ \exists \beta_X$ respectively correspond to the standard interpretations of the modalities $\Box$ and $\Diamond$ in the coalgebra $(X,\alpha_X)$.

\medskip
Moss' classical Nabla operator~\cite{Moss99} can also be defined as a family $\nabla$ of $n$-ary quantifiers $\nabla^n : \Sub^n \Rightarrow \Sub$ defined for every coalgebra $(X,\alpha_X)$. Let us suppose a natural transformation $\beta : F \Rightarrow \cP$. Then, we can define internally the natural transformation $\lambda : \cP \circ \cP \Rightarrow \cP \circ F^{op}$ as follows: let $X \in |\cC|$ be an object and let $Y:PPX$ be a variable

$$\lambda_X(Y) = 
\left\{
    \begin{array}{l|c}
         & (\forall x, x \in_{PX} Y \Rightarrow (\exists z, z \in_X \beta_X(y) \wedge z \in_X x)) \\
   y:F(X)  & \wedge \\
         & (\forall z, z \in_X \beta_X(y) \Rightarrow (\exists x,  x \in_{PX} X \wedge z \in_X x))
    \end{array}
\right\}
$$

Then, for every $n \in \mathbb{N}$, let us set $\nabla^n_X(x_1,\ldots,x_n) = \alpha^{-1}_X(\lambda_X(\{x_1,\ldots,x_n\}))$. $\nabla^n$ is a natural transformation as $\alpha^{-1}$ and $\lambda$ are. Likewise, we can observe that the following formula expressed in the internal language of the topos $\cC$
$$x_1 \preceq_X y_1 \wedge \ldots \wedge x_n \preceq_X y_n \Rightarrow \lambda_X(\{x_1,\ldots,x_n\}) \preceq_{PX} \lambda_X(\{y_1,\ldots,y_n\})$$
is satisfied, ensuring that $\nabla^n_X$ is isotone. 

Before presenting the internal logic of~\cite{AB23}, we point out that, quite interestingly, propositional connectives are subsumed by our definition of quantifiers.

\begin{remark}
The propositional operators $\wedge$ and $\vee$ can be defined as quantifiers of arity $2$ from the context identity $Id_\sigma : \sigma \to \sigma \in \Ctx$:
\[
    \wedge : Prop_{\cC}(|\cM|(\sigma)) \to Prop_{\cC}(|\cM|(\sigma));(\iota,\iota') \mapsto \iota \wedge \iota'
\]
and 
\[
    \vee : Prop_{\cC}(|\cM|(\sigma)) \to Prop_{\cC}(|\cM|(\sigma));(\iota,\iota') \mapsto \iota \vee \iota'
\]
Similarly, extending the definition of quantifiers to either isotone or antitone mappings would enable the definition of negation as a quantifier of arity \(1\). 
\end{remark}

\subsection{Internal logic}
\label{subsec:internallogic}

\subsubsection{Syntax}

Standardly, formulas are defined inductively from basic formulas. Basic formulas are usually directly interpretable in models. This leads to the following definition. As is customary, we assume that each quantifier $\cQ f$ of arity $n$ has its syntactic equivalent, also denoted $\cQ f$, used to construct formulas.

\begin{definition}[Basic formulas]
\label{def:basic formulas}
Let $\Ctx$ be a category of contexts. A set of {\bf basic formulas} is a $|\Ctx|$-indexed family of sets $(Bc_\sigma)_{\sigma \in |\Ctx|}$. \\
$Bc$ is said {\bf interpretable} in a semantical system $\cS = (\cC,\Ctx,Mod,|\_|)$ if it is equipped for every model $\cM \in |Mod|$ and every context $\sigma \in |\Ctx|$ with a mapping $\sem{\cM}_\sigma(\_) : Bc_\sigma \to Prop_\cC(|\cM|(\sigma))$ satisfying the following property: for every family of models $(\cM_i)_{i \in I}$, %
and for every $\delta_I \in Prop_\cC(|\prod_I \cM_i|(\sigma))$, 
\begin{center}
\(\delta_I \preceq_{|\prod_I \cM_i|(\sigma)} \sem{\prod_I \cM_i}(\sigma.bc)\) iff for all $i \in I$, $\delta_I \preceq_{|\prod_I \cM_i|(\sigma)} |p_{I,i}|^*_\sigma(\sem{\cM_i}(\sigma.bc))$
\end{center}
\end{definition}

The careful reader will have noticed that we are surcharging the notation \(\sem{\cM}_{[\vec{x}]}(\_)\) used in Section~\ref{fol-semanticalsystem} for the interpretation of terms. Indeed, the mapping $\sem{\cM}_\sigma(\_) : Bc_\sigma \to Prop_\cC(|\cM|(\sigma))$ will later be extended to interpret formulas. In FOL, this extension will build up on the interpretation of terms already introduced.

\begin{example}[Basic formulas in first-order categorical logic]
\label{ex:basic formulas for FOL}
Let $\Sigma = (S,F,R)$ be a signature. The set of basic formulas for the first-order logic is the standard set of atomic formulas. More formally, we define, for every context $[\vec{x}]$, the set $Bc_{[\vec{x}]}$ by:
\[Bc_{[\vec{x}]} = 
\left\{
\begin{array}{c}
\{r(t_1,\ldots,t_n) \mid \vec{x}~\text{suitable context for}~t_i\} \\
\cup \\
\{t = t' \mid \vec{x}~\text{suitable context for $t$ and $t'$}\}
\end{array}
\right.\]
where a context $\vec{x}$ is suitable for a term or a formula if each free variable of this term or this formula occurs in $\vec{x}$.

Let $\cS = (\cC,\Ctx,Mod,|\_|)$ be the semantical system of Section~\ref{fol-semanticalsystem}. Given a model $\cM$, we define the satisfaction mapping as $\sem{\cM}_{[\vec{x}]}(\_) : Bc_{[\vec{x}]} \to \Sub(|\cM|([\vec{x}]))$ such that 
\begin{itemize}
\item for any equation $t = t'$, $\sem{\cM}_{[\vec{x}]}(t = t')$ is the equalizer of
    \[\begin{tikzcd}[column sep=5em]
    	{|\cM|([\vec{x}])} & {M_s}
    	\arrow["{\sem{\cM}_{[\vec{x}]}(t')}"', shift right, from=1-1, to=1-2]
    	\arrow["{\sem{\cM}_{[\vec{x}]}(t)}", shift left, from=1-1, to=1-2]
    \end{tikzcd}\]
where \(s\) is the common sort of \(t\) and \(t'\);
\item for any relation $r(\vec{t})$, $\sem{\cM}_{[\vec{x}]}(r(\vec{t}))$ is the subobject $O' \mto |\cM|([\vec{x}])$ given by the pullback:
    \[\begin{tikzcd}[column sep=5em]
    	{O'} & O \\
    	{|\cM|([\vec{x}])} & {M_{s_1} \times \ldots \times M_{s_n}}
    	\arrow[from=1-1, to=1-2]
    	\arrow["{r^\cM}", tail, from=1-2, to=2-2]
    	\arrow["{\sem{\cM}_{[\vec{x}]}(r(\vec{t}))}"', tail, from=1-1, to=2-1]
    	\arrow["{\sem{\cM}_{[\vec{x}]}(\vec{t})}"', from=2-1, to=2-2]
    \end{tikzcd}\]
if $r : s_1 \times \ldots \times s_n$. 
\end{itemize}

The condition is obviously satisfied because, by the definition of model products, we have that 
\[\sem{\prod_I \cM_i}_{[\vec{x}]}(bc) = \prod_I \sem{\cM_i}_{[\vec{x}]}(bc)\] 
\end{example}

\begin{example}[Basic formulas in higher-order categorical logic]
\label{ex:basic formulas for HOL}
Let $\Sigma = (S,F,R)$ be a signature. The set of basic formulas for the higher-order logic is the set of atomic formulas defined for every context $[\vec{x}]$ by:

\[Bc_{[\vec{x}]} = 
\left\{
\begin{array}{c}
\{r(t_1,\ldots,t_n) \mid \vec{x}~\mbox{suitable context for}~t_i\} \\
\cup \\
\{t = t' \mid \vec{x}~\mbox{suitable context for $t$ and $t'$}\} \\
\cup \\
\{t \in_A t' \mid t:A, t':PA, \vec{x}~\mbox{suitable context for $t$ and $t'$}\}
\end{array}
\right.\] 

Given a model $\cM$, equations and predicates are satisfied as in FOL. $\sem{\cM}_{[\vec{x}]}(t \in_A t')$ is the subobject $O \mto |\cM|([\vec{x}])$ such that $O$ is the pullback of the diagram

\[\begin{tikzcd}[column sep=10em]
	O & {\in_A} \\
	{|\cM|([\vec{x}])} & {M_A \times PM_A}
	\arrow[from=1-1, to=1-2]
	\arrow[tail, from=1-2, to=2-2]
	\arrow["{\sem{\cM}_{[\vec{x}]}(t \in_A t')}"', tail, from=1-1, to=2-1]
	\arrow["{(\sem{\cM}_{[\vec{x}]}(t), \sem{\cM}_{[\vec{x}]}(t'))}"', from=2-1, to=2-2]
\end{tikzcd}\]

Hence, for every $A \in \Sigma\mbox{-Typ}$, $\in_A$ can be seen as a new relation name with profile $A \times PA$. Then, the property of Definition~\ref{def:basic formulas} follows from the same arguments as in Example~\ref{ex:basic formulas for FOL}.
\end{example}

\begin{example}[Basic formulas in institutions]
\label{ex:basic formulas in Institutions}
In institutions, formulas being simple elements of a set, the notion of atomic formulas can only be semantically approximated. Hence, in institutions, atomic formulas are not explicitly considered as such but rather implicitly from their model-theoretic properties. Hence, given a signature $\Sigma \in |Sig|$, a subset of formulas $E \subseteq Sen(\Sigma)$ is said {\bf basic}~\cite[Chap. 5, p. 108]{Dia08} if there exists a model $\mathcal{M}_E \in |Mod(\Sigma)|$ such that for each model $\mathcal{M} \in |Mod(\Sigma)|$
\[
\mathcal{M} \models_\Sigma E \; \mbox{iff} \; \mbox{there exists a morphism}~ \mu : \mathcal{M}_E \to \mathcal{M}
\]

Let $\mathcal{S}_\mathcal{I}(\Sigma)$ be the semantical system defined in Section~\ref{subsubsec:semantical system for institutions}.
For a signature \(\Sigma'\) and an object \(\chi : \Sigma \to \Sigma'\) in \(|\Ctx|\), we consider a subset $Bc_\chi$ of $Sen(\Sigma')$ to be basic whenever each formula $bc \in Bc_\chi$ is basic in the previous meaning, i.e., if there exists a model $\cM_{bc} \in |Mod(\Sigma')|$ such that for each model $\cM' \in |Mod(\Sigma')|$
\[
\cM' \models_{\Sigma'} bc \; \mbox{iff} \; \mbox{there exists a morphism}~ \mu : \cM_{bc} \to \cM'
\]
Then, for $\cM \in Mod(\Sigma)$ and $bc \in Bc_\chi$
\[
    \sem{\cM}_\chi(bc) = \{\cM' \in |\cM|(\chi) \mid \cM' \models_{\Sigma'} bc\}
\]

Let us show that \((Bc_{\chi})_{\chi \in |Ctx|}\) is interpretable in the semantical system of institution (see Section~\ref{subsubsec:semantical system for institutions}).
Let $(\cM_i)_{i \in I}$ be a family of models in $Mod(\Sigma)$. Let $\chi : \Sigma \to \Sigma' \in |\Ctx|$. As $Mod(\chi)$ creates small products, for every basic formula $bc \in Bc_\chi$, we can write:
\[
    \sem{\prod_I \cM_i}_\chi(bc)  = \{\prod_I \cM'_i \in |\prod_I \cM_i|(\chi) \mid \prod_I \cM'_i \models_{\Sigma'} bc\}
\]

Let $S$ be a subset of $\sem{\prod_I \cM_i}_\chi(bc)$. 
If we have for every $i \in I$ that there exists a subset $S_i$ of $\sem{\cM_i}_\chi(bc)$ such that $S \subseteq |p_{I,i}|^*_\chi(S_i)$, then this means that for every $\prod_I \cM'_i \in S$, $\cM'_i \in S_i$ for all $i \in I$. Hence, there exists a model $\cM_{bc}$ and a morphism $\mu_i : \cM_{bc} \to \cM'_i$ for all $i \in I$. By the universal property of small products, there is unique morphism $\mu : \cM_{bc} \to \prod_I \cM'_i$ from which we can conclude that $\prod_I \cM'_i \models_{\Sigma'} bc$.
\end{example}

\begin{example}[Basic formulas in modal logic]
\label{ex:basic formula in modal logic}
In Section~\ref{modal-semanticalsystem}, coalgebras have been defined over the empty propositional signature. Hence, the set of basic formulas is empty. Within the set framework, predicate liftings generalize atomic propositions. Indeed, for $PV$ a set of propositional variables, we can consider the functor \[F : \Set \to \Set;S \mapsto \powerset(S) \times \powerset(PV)\] Clearly, $F$-coalgebras $(S,\alpha)$ give rise to Kripke models. Then, given a propositional variable $p \in PV$, by considering the predicate lifting 
\[\lambda_p : S \mapsto \{\alpha(s) \mid s \in S, p \in \pi_2(\alpha(s))\}\] 
where $\pi_2 : F(S) \to \powerset(PV)$ is the second projection, we have that $p$ is the formula $[\lambda_p](\top) = \alpha^{-1}_p (\lambda_p(\top))$. 

In a more general framework (i.e., when the category $\cC$ of the functor $F$ is not necessarily $\Set$), given a set of propositional variables $PV$, models have to be defined as pairs $((X,\alpha_X),\nu)$ where $(X,\alpha_X)$ is a coalgebra and $\nu : PV \to \Sub(X)$ is a mapping. Morphisms between co-algebras are extended by imposing that $\nu(p) \preceq_X |\mu|^*(\nu'(p))$ for $\mu : ((X,\alpha_X),\nu) \to ((Y,\alpha_Y),\nu')$. In this case, we have that $\sem{((X,\alpha_X),\nu)}_\bullet(p) = \nu(p)$.

The property of Definition~\ref{def:basic formulas} is even more easily demonstrated from the same arguments as in Example~\ref{ex:basic formulas for FOL}.
\end{example}

We now have all the ingredients to formally define our notion of abstract logic.

\begin{definition}[Abstract categorical logic]
\label{def:abstract logic}
A {\bf logic} is given by a tuple $\cL = (\cS,\cQ,Bc)$ where
\begin{itemize}
	\item $\cS$ is semantical system;
	\item $Q = (\cQ_n)_{n \in \mathbb{N}}$ is a $\mathbb{N}$-indexed family of sets of quantifiers, i.e., for every $n \in \mathbb{N}$, $\cQ_n$ is a set of $n$-ary quantifiers;
	\item $Bc$ is a set of basic formulas interpretable in $\cS$.
\end{itemize}
\end{definition}

\begin{example}[Intuitionistic propositional logic (IPL)]
\label{ex:intuitionistic propositional logic}
Let $PV$ be a set of propositional variables. Since intuitionistic propositional logic is sound with respect to Kripke semantics, where the relation is a preorder, it is natural to define the logic for intuitionistic reasoning as:
\begin{itemize}
    \item the semantical system $(\cC,\Ctx,Mod,|\_|)$ where:
    \begin{itemize}
        \item $\Ctx$ is the trivial category where the unique object is $\bullet$;
        \item $\cC$ is a topos to which we associate the contravariant functor $\Sub$;
        \item $Mod$ is the category of pairs $(\alpha,\nu)$ where $\alpha = (X,b:X \to PX)$ is a coalgebra and $\nu : PV \to \Sub(X)$ is a mapping (see Example~\ref{ex:basic formula in modal logic}) such that the underlying relation $R_b$ is a preorder, i.e., the morphism $b$ satisfies the two following formulas expressed in the internal language of the topos $\cC$:        
        \[
            \forall x,x \in_X b(x)
        \]
        \[
            \forall x,\forall y,\forall z, y \in_X b(x) \wedge z \in_X b(y) \Rightarrow z \in_X b(x)
        \]
        \item $|\_|$ is the functor defined in Section~\ref{modal-semanticalsystem}.
    \end{itemize}
    \item for every $n \in \mathbb{N}$, $\mathcal{Q}_n = \emptyset$;
    \item $Bc = PV$.
\end{itemize}
\end{example}

\begin{example}[Fist-order logic (FOL)]
\label{ex:FOL logic}
Let $\Sigma$ be a multi-sorted FOL signature. Let $V$ be a set of sorted variables. The logic for FOL is defined by:
\begin{itemize}
    \item $\cS_\Sigma$ is the semantical system presented in Section~\ref{fol-semanticalsystem};
    \item for every $n \neq 1$, $\cQ_n = \emptyset$, and $\cQ_1 = \{\forall \gamma,\exists \gamma \mid \gamma~\mbox{is a morphism in}~\Ctx\}$;
    \item $Bc$ is the set of basic formulas presented in Example~\ref{ex:basic formulas for FOL}.
\end{itemize}
\end{example}

\begin{example}[Higher-order logic (HOL)]
Let $\Sigma$ be a multi-sorted FOL signature. Let $V$ be a set of typed variables. The logic for HOL is defined by:
\begin{itemize}
    \item $\cS_\Sigma$ is the semantical system presented in Section~\ref{hol-semanticalsystem};
    \item for every $n \neq 1$, $\cQ_n = \emptyset$, and $\cQ_1 = \{\forall \gamma,\exists \gamma \mid \gamma~\mbox{is a morphism in}~\Ctx\}$;
    \item $Bc$ is the set of basic formulas presented in Example~\ref{ex:basic formulas for HOL}.
\end{itemize}
\end{example}

\begin{example}[Institutional FOL]
\label{ex:institutional FOL}
Let $\mathcal{I} = (Sig,Sen,Mod,\models)$ be an institution such that the functor $Sen$ has a subfunctor $Bc : Sig \to \Set$ such that for every $\Sigma \in |Sig|$, every $bc \in Bc(\Sigma)$ is basic. Let $\Sigma \in |Sig|$ be a signature. The institutional FOL for \(\Sigma\) is defined by:
\begin{itemize}
    \item $\mathcal{S}_\mathcal{I}(\Sigma)$ is the semantical system presented in Section~\ref{subsubsec:semantical system for institutions};
    \item for every $n \neq 1$, $\mathcal{Q}_n = \emptyset$ and $\mathcal{Q}_1 = \{\forall \theta,\exists \theta \mid \theta : \chi \to \chi' \in \Ctx\}$;
    \item for every signature morphism $\chi : \Sigma \to \Sigma'$, $Bc_\chi = Bc(\Sigma')$.
\end{itemize}
\end{example}

\begin{example}[Coagebraic modal logic]
\label{ex:coalgebraic modal logic}
Let $\Lambda = (\Lambda_n)_{n \in \omega}$ be a family of sets of $n$-ary predicate liftings. Let $F : \cC \to \cC$ be a functor where $\cC$ is a topos. Coalgebraic modal logic is then defined by:
\begin{itemize}
    \item $\cS_F$ is the semantical system presented in Section~\ref{modal-semanticalsystem};
    \item For every $n \in \mathbb{N}$, $\cQ_n = \{[\lambda] \mid \lambda \in \Lambda_n\}$;
    \item $Bc = PV$ where $PV$ is a set of propositional variables and then models are tuples of the form $(\alpha,\nu)$ where $\alpha = (X,\alpha_X)$ is a coalgebra and $\nu : PV \to \Sub(X)$ is a mapping.
\end{itemize}
\end{example}

\begin{example}[Moss' colagebraic logic]
Let $F : \cC \to \cC$ where $\cC$ is a topos. Moss' coalgebraic modal logic is defined by:
\begin{itemize}
    \item $\mathcal{S}_F$ is the semantical system presented in Section~\ref{modal-semanticalsystem};
    \item For every $n \in \mathbb{N}$, $\mathcal{Q}_n = \{\nabla^n\}$ (assuming a natural transformation $\beta : F \to \mathcal{P}$ as in Section~\ref{modal:quantifiers});
    \item $Bc = PV$ with $PV$ a set of propositional variables.
\end{itemize}
\end{example}

The general set of formulas can now be defined inductively over the logical connectives and quantifiers.

\begin{definition}[Formulas]
\label{def:formulas}
Let $\cL = (\cS,\cQ,Bc)$ be a logic. The set $\cF_\cL$ of {\bf formulas} for \(\cL\) is defined inductively as follows:

\begin{itemize}
	\item for every $bc \in Bc_\sigma$, $\sigma.bc \in \cF_\cL$;
	\item for every $\sigma \in |\Ctx|$, $\sigma.\bot \in \cF_\cL,\sigma.\top \in \cF_\cL$;
	\item if $\sigma.\varphi \in \cF_\cL$ and $\sigma.\psi \in \cF_\cL$, then $\sigma.(\varphi \wedge \psi) \in \cF_\cL, \sigma.(\varphi \vee \psi) \in \cF_\cL, \sigma.(\varphi \Rightarrow \psi) \in \cF_\cL$; 
	\item if $\sigma.\varphi \in \cF_\cL$, then $\sigma.\neg \varphi \in \cF_\cL$;
	\item for every $f : \sigma \to \tau \in \Ctx$:
		\begin{itemize}
			\item if $\sigma.\varphi_1 \in \cF_\cL$,$\ldots$, $\sigma.\varphi_n \in \cF_\cL$, then $\tau.\Qf (\varphi_1,\ldots,\varphi_n) \in \cF_\cL$, where $\Qf$ is a $n$-ary quantifier name whose semantics is defined by an isotone mapping $\Qf$ as introduced in Definition~\ref{def:quantifier};
			\item if $\tau.\varphi \in \cF_\cL$, then $\sigma.f(\varphi) \in \cF_\cL$.
		\end{itemize}
\end{itemize}
\end{definition}

\subsubsection{Semantics}

The internal logic is defined as an extension of PL by adding quantifiers. 

\begin{definition}[Formula interpretation]
\label{def:formula interpretation}
Let $\cL = (\cS,\cQ,Bc)$ be a logic. Let $\cM \in |Mod|$ be a model. We define the mapping 
\[\sem{\cM} : \cF_\cL \to \bigcup_{\sigma \in |\Ctx|}Prop_\cC(|\cM|(\sigma))\] 
such that for every $\sigma.\varphi \in \cF_\cL$, $\sem{\cM}(\sigma.\varphi) \in Prop_\cC(|\cM|(\sigma))$, as the canonical extension of $(\sem{\cM}_\sigma(\_))_{\sigma \in |\Ctx|}$ to formulas in $\cF_\cL$ as follows:

\begin{itemize}
\item $\sem{\cM}(\sigma.\bot) = \bot_{|\cM|(\sigma)}$ and $\sem{\cM}(\sigma.\top) = \top_{|\cM|(\sigma)}$;
\item for every $bc \in Bc_\sigma$, $\sem{\cM}(\sigma.bc) = \sem{\cM}_\sigma(bc)$;
\item $\sem{\cM}(\sigma.(\varphi \wedge \psi)) = \sem{\cM}(\sigma.\varphi) \wedge \sem{\cM}(\sigma.\psi)$; 
\item $\sem{\cM}(\sigma.(\varphi \vee \psi)) = \sem{\cM}(\sigma.\varphi) \vee \sem{\cM}(\sigma.\psi)$;
\item $\sem{\cM}(\sigma.\neg \varphi) = \sem{\cM}(\sigma.\varphi) \to \sem{\cM}(\sigma.\bot)$;\footnote{The arrow symbol for the interpretation of \(\sem{\cM}(\sigma.\neg \varphi)\) and \(\sem{\cM}(\sigma.\varphi \Rightarrow \psi)\) is the implication in the Heyting algebra \(Prop_\cC(|\cM|(\sigma))\).}
\item $\sem{\cM}(\sigma.\varphi \Rightarrow \psi) = \sem{\cM}(\sigma.\varphi) \to \sem{\cM}(\sigma.\psi)$;
\item $\sem{\cM}(\tau.\Qf(\varphi_1,\ldots,\varphi_n)) = \Qf_\cM(\sem{\cM}(\sigma.\varphi_1),\ldots,\sem{\cM}(\sigma.\varphi_n))$ with $f : \sigma \to \tau \in \Ctx$.
\item $\sem{\cM}(\sigma.f(\varphi)) = |\cM|(f)^*(\sem{\cM}(\tau.\varphi))$ with $f : \sigma \to \tau \in \Ctx$.
\end{itemize} 
\end{definition}

In first-order and higher-order logics, formulas of the $\sigma.f(\varphi)$ are the counterpart of variable substitutions. 

\begin{proposition}[\cite{AB23}]
\label{prop:counterpart substitution}
In FOL and HOL, for every morphism $\gamma = [t_1:s_1,\ldots,t_n:s_n] : [\vec{x}] \to [\vec{y}]$ with $\vec{y} = (y_1:s_1,\ldots,y_n:s_n)$, every formula $\varphi$ such that the variables of $\varphi$ occurs in $\vec{y}$, and every model $\cM \in |Mod|$, the following equality holds:
\[\sem{\cM}([\vec{x}].\gamma(\varphi)) = \sem{\cM}([\vec{x}].\varphi(y_1/t_1,\ldots,y_n/t_n))\] 
where $\varphi(y_1/t_1,\ldots,y_n/t_n)$ is the formula $\psi$ obtained from $\varphi$ by substituting all free occurrences of $y_i$ by $t_i$. 
\end{proposition}

We recall here the main steps of the proof given in~\cite{AB23} for the sake of completeness.

\begin{proof}
The proof is done by structural induction on $\varphi$. The proof is straightforward for general cases. The only difficulties may come from basic cases. So let us prove the statement for formulas of the form $t = t'$ and $r(\vec{t})$ (the case $t \in_A t'$ is treated in a similar way to $r(\vec{t})$).

\medskip
Let $\varphi$ be of the form $t = t'$ with $t,t' : [\vec{y}] \to [\vec{z}]$. From Example~\ref{ex:basic formulas for FOL}, $\sem{\mathcal{M}}([\vec{y}].t = t')$ equalizes $\sem{\mathcal{M}}_{[\vec{y}]}(t)$ and $\sem{\mathcal{M}}_{[\vec{y}]}(t')$.

By structural induction on terms, we have  
\[\sem{\mathcal{M}}_{[\vec{x}]}(t(\vec{y}/\gamma)) = \sem{\mathcal{M}}_{[\vec{y}]}(t) \circ (\sem{\mathcal{M}}_{[\vec{x}]}(t_1),\ldots,\sem{\mathcal{M}}_{[\vec{x}]}(t_n))\]
where $\sem{\mathcal{M}}_{[\vec{x}]}(t(\vec{y}/\gamma))$ is a shorter notation for $\sem{\mathcal{M}}_{[\vec{x}]}(y_1/t_1,\ldots,y_n/t_n)$.

\medskip
Thus, $\sem{\mathcal{M}}([\vec{x}].t(\vec{y}/\gamma) = t'(\vec{y}/\gamma))$ equalizes $\sem{\mathcal{M}}_{[\vec{x}]}(t(\vec{y}/\gamma))$ and $\sem{\mathcal{M}}_{[\vec{x}]}(t'(\vec{y}/\gamma))$. Hence, by definition of the pullback functor $|\mathcal{M}|(\gamma)^*$ and the universal property of equalizers, we necessarily have that 
\[
    |\mathcal{M}|(\gamma)^*(\left[\sem{\mathcal{M}}([\vec{y}].t = t') \mto |\mathcal{M}|([\vec{y}])\right])
    = \left[\sem{\mathcal{M}}([\vec{x}].t(\vec{y}/\gamma) = t'(\vec{y}/\gamma)) \mto |\mathcal{M}|([\vec{x}])\right]
\]

Let $\varphi$ be of the form $r(\vec{t})$. Then, by the definition of both semantics of $r(\vec{t})$ and of pullback functor, the following diagram 
\[\begin{tikzcd}[column sep=5em]
	{O'} & O & {r^\cM} \\
	{|\cM|([\vec{x}])} & {|\cM|([\vec{y}])} & {M_r}
	\arrow[from=1-1, to=1-2]
	\arrow[from=1-2, to=1-3]
	\arrow[tail, from=1-3, to=2-3]
	\arrow["{|\cM|(\gamma)^*(\iota)}"', tail, from=1-1, to=2-1]
	\arrow["{\sem{\cM}_{[\vec{x}]}(\gamma)}"', from=2-1, to=2-2]
	\arrow["{\sem{\cM}_{[\vec{y}]}(t)}"', from=2-2, to=2-3]
	\arrow["\iota"', tail, from=1-2, to=2-2]
\end{tikzcd}\]
commutes.
\end{proof}

\begin{definition}[Validation]
\label{def:validation}
Let $\cL = (\cS,\cQ,Bc)$ be a logic. Given a model $\cM \in |Mod|$ and a formula $\sigma.\varphi \in \cF_\cL$, we write $\cM \models \sigma.\varphi$ if $\sem{\cM}(\sigma.\varphi) = \top_{|\cM|(\sigma)}$. Moreover, for every $\iota \in Prop_\cC(|\cM|(\sigma))$, we write $\cM \models_\iota \sigma.\varphi$ if $\iota \preceq_{|\cM|(\sigma)} \sem{\cM}(\sigma.\varphi)$. 
\end{definition}

\section{Ultraproducts and \Los's theorem}
\label{sec:ultraproduct}

In this section, we prove our main contribution, namely the ultraproduct method and its fundamental theorem in abstract categorical logic. To obtain Theorem~\ref{th:Los}, we need filtered products (see Section~\ref{sec:filteredproducts}) along with some further requirements on the various categories. More precisely, we introduce additional requirements on the semantical system to handle intersections in the filter (sets of generators and the finiteness condition) and disjunctions of formulas (coverage condition) in Subsection~\ref{subsec:los:systems}. Similar conditions are added to handle quantifiers in Subsection~\ref{subsec:los:quantifiers} and pullback functors in Subsection~\ref{subsec:los:pullbacks}.
The main result (\Los's theorem) is proved in Subsection~\ref{subsec:los:theorem}.
We finally discuss dual quantifiers in Subsection~\ref{subsec:los:duality} and the standard corollary of \Los's theorem, namely the compactness theorem, in Subsection~\ref{subsec:los:compactness}.

In the following, we consider a logic \(\cL = (\cS,\cQ,Bc)\) with the semantical system \(\cS = (\cC,\Ctx,Mod,|\_|)\) and the interpretation of formulas given by the mapping \(\sem{\_}\) as in Definition~\ref{def:formula interpretation}.

\subsection{Filtered semantical systems}
\label{subsec:los:systems}
So far, our abstract categorical logic has been presented as generically as possible. 
Indeed, our semantical systems link a prop-category - expected to provide the needed structure - and a category of models. Still, we need conditions to retrieve the structure in the model category from the prop-category. Here, the structure is to have filtered products and thus ultraproducts on families of models, which is needed to obtain an abstract version of \Los's theorem.
We now provide conditions to be able to properly consider ultraproducts on families of models when dealing with our abstract categorical logic. We first discuss assumptions on the semantical systems.

\subsubsection{Sup-generation}

In the context of presheaves, Proposition~\ref{prop:filtered product in presheaves} highlights that filtered products (as would any colimit) are computed componentwise. However, this construction does not hold in arbitrary prop-categories. Locally finitely presentable categories (see Definition~\ref{def:lfp category}) are intrinsically endowed with such a construction by considering a set of generators; similarly, the functor \(\Sub\) in toposes simplifies the reasoning about intersections. In an arbitrary prop-category, we can only impose additional conditions on the structure carrying the ingredients for building the formulas, i.e., the Heyting algebra. In a sense, we solve the difficulty of obtaining filtered products the same way as in locally finitely presentable categories: via a set of generators.

\begin{definition}[Sup-generator~\cite{serra_connectivity_1998}]
\label{def:supgenerator}
    Let \(L\) be a lattice. A subset \(X \subseteq L\) is a {\bf sup-generator} of \(L\) when any element \(a \in L\) is the supremum of the elements of \(X\) that it majorates: 
    \[
    \forall a \in L, a = \bigvee \downarrow\!a
    \]
    where $\downarrow\!a = \{x \in X \mid x \preceq_L a\}$ for $a \in L$.
    Then \(L\) is said to be {\bf sup-generated} by \(X\).
\end{definition}

We always consider downward closed sup-generators, i.e., that for any element \(a\) of a lattice \(L\) sup-generated by \(X\), if there exists \(x \in X\) such that \(a \preceq_L x\), then \(a \in X\). Note that if \(L\) is sup-generated by \(X\), then \(L\) is sup-generated by the downward closure of \(X\) in \(L\). 

\begin{definition}[Sup-generated prop-category]
\label{def:is supgenerated}
    A prop-category $\cC$ is {\bf sup-generated} if \(Prop_\cC(X)\) is sup-generated for each \(X\) in \(|\cC|\) by a set \(l_X \setminus \{\bot_X\}\)
    such that for all \(f:X \to Y\) in \(\cC\), for all \(\delta_X\) in \(l_X\), and for all \(\iota\) in \(Prop_\cC(Y)\), if \(\delta_X \preceq_X f^*(\iota)\), then there exists \(\delta_Y\) in \(l_Y\) such that \(\delta_Y \preceq_Y \iota\) and \(\delta_X \preceq_X f^*(\delta_Y)\).
\end{definition}

\begin{remark}
In the context of topos, this notion of generators refines that of global elements. In classical terms, a global element in a category $\mathcal{C}$ endowed with a terminal object $\mathbb{1}$ is a morphism $\mathbb{1} \to X$, thereby extending the familiar idea of points in the category of sets. Yet, this notion of global elements does not allow identifying an object `points' in an arbitrary category. For instance, consider the category of graphs, where the terminal object is a graph containing a single vertex with a loop (an edge connecting the vertex to itself). Then, the global elements of a graph are restricted to vertices with a loop. A more permissive viewpoint classifies the elements of a graph as isolated vertices, vertices with loops, and pairs of vertices connected by an edge.
\end{remark}

From now on, $l_X$ denotes the set of generators of $X$. %

\begin{example}[Atomic toposes~\cite{Car12}]
Let $\mathcal{C}$ be a topos. An object $X \in |\mathcal{C}|$ is an \emph{atom} if its only subobjects are $Id_X$ and $\emptyset \to X$ (up to isomorphism), and they are distinct from each other. $\mathcal{C}$ is an \emph{atomic topos} if it is an elementary topos that possesses an atomic geometric morphism $\mathcal{C} \to \Set$ (i.e., its inverse image functor is logical). If $\mathcal{C}$ is further a Grothendick topos,\footnote{i.e., $\mathcal{C}$ is equivalent to the category of sheaves on a site~\cite{Car17}.} the subobject lattice of every object of $X \in |\mathcal{C}|$ is a complete atomic Boolean algebra. Then, every object can be written as disjoint unions of atoms. 
\end{example}

\begin{example}[Locally finitely presentable categories]
\label{ex:supgeneretion:presheaves}
Not all toposes are atomic. For instance, it is easy to see that the subobject lattice of a graph is only a Heyting algebra and not a Boolean algebra and that, in particular, a graph is not a disjoint union of vertices without arcs (which are the atomic objects in the category of graphs).
Recall that in the case of locally finitely presentable categories (see Definition~\ref{def:lfp category}), any object is a filtered colimit of the canonical diagram of finitely presentable objects mapping into it. Thus, the `good properties' of the functor \(\Sub\) ensure that any locally finitely presentable topos is a sup-generated prop-category. Locally finitely presentable toposes encompass presheaves and coherent toposes~\cite[Section D.3.3]{Johnstone02}.

\end{example}

\begin{example}[Institutions]
Let \(\cM \in |Mod(\Sigma)|\) be a model and \(\chi : \Sigma \to \Sigma'\) a context in an institution, i.e., for the semantical system of Section~\ref{subsubsec:semantical system for institutions}. Then $Prop(|\cM|(\chi))$ is a power set, meaning that it is sup-generated by singletons, i.e., any set $\{\cM'\}$ such that $\cM' \in |Mod(\Sigma')|$ and $Mod(\chi)(\cM') = \cM$.  
\end{example}

\begin{proposition}
    If $\cC$ is sup-generated, then for all models \(\cM \in |Mod|\), for all formulas \(\sigma.\varphi \in \cF_\cL\), for all \(\iota \in Prop_\cC(|\cM|(\sigma))\), \(\cM \models_\iota \sigma.\varphi\) if and only if for all \(\iota' \in l_{|\cM|(\sigma)}\), \(\iota' \preceq_{|\cM|(\sigma)} \iota\) implies \(\cM \models_{\iota'} \sigma.\varphi\).
\end{proposition}

\begin{proof}
    Let \(\cC\) be a sup-generated prop-category, \(\cM \in |Mod|\) be a model, \(\sigma.\varphi\) be a formula, and \(\iota\) in \(Prop_\cC(|\cM|(\sigma))\).
    Since \(\cC\) is sup-generated, \(\iota = \bigvee \downarrow\!\iota\).

    If \(\cM \models_\iota \sigma.\varphi\), then \(\iota \preceq_{|\cM|(\sigma)} \sem{\cM}(\sigma.\varphi)\). In particular, for all \(\iota' \in l_{|\cM|(\sigma)}\) such that \(\iota' \preceq_{|\cM|(\sigma)} \iota\), it holds that \(\iota' \preceq_{|\cM|(\sigma)} \sem{\cM}(\sigma.\varphi)\), i.e., \(\cM \models_{\iota'} \sigma.\varphi\).
    The reverse implication holds similarly by unfolding the definition of \(\iota\) as a supremum.
\end{proof}

\subsubsection{Coverage}

Sup-generation ensures that we can deal with intersections by reasoning on generators. Similarly, we introduce a notion of coverage to handle disjunctions.

\begin{definition}[Covered prop-category]
    A prop-category $\cC$ is {\bf covered} if for every $\iota \in Prop_\cC(X)$ and every $\delta_1,\delta_2 \in Prop_\cC(X)$, if $\iota \preceq_X \delta_1 \vee \delta_2$, then there exists $\iota_1,\iota_2 \in Prop_\cC(X)$ such that $\iota = \iota_1 \vee \iota_2$ and $\iota_i \preceq_X \delta_i$ for \(i=1,2\).
\end{definition}

\begin{proposition}
\label{prop:covering property}
Elementary toposes are covered.
\end{proposition}

\begin{proof}
If $Z \preceq_Y X_1 \vee X_2$, then it is sufficient to define each \(Z_i\) as $Z_i = Z \wedge X_i$ for $i = 1,2$.

\end{proof}

\subsubsection{Filtered semantical systems}
\begin{definition}[Filtered semantical systems]
\label{def:filteredsemsyst}
    A semantical system $\cS = (\cC,\Ctx,Mod,|\_|)$ is said to be {\bf filtered} if it satisfies the following properties:

    \begin{description}
        \item[Sup-generation:] $\cC$ is sup-generated.
        \item[Covering property:] \(\cC\) is covered;
        \item[Filtered models:] $Mod$ has filtered products. 
        \item[Projection:] for every context $\sigma \in |\Ctx|$, every family of models $(\cM_i)_{i \in I}$ and every family $(\iota_i)_{i \in I}$ where for every $i \in I$, $\iota_i \in Prop_\cC(|\cM_i|(\sigma))$, there exists $\iota_I \in Prop_\cC(|\prod_I \cM_i|(\sigma))$ such that $\downarrow\!\iota_I = \bigcap_{i \in I} \downarrow\!|p_{I,i}|^*_\sigma(\iota_i)$.
        \item[Finiteness:] for every model $\cM \in |Mod|$, every context \(\sigma \in |\Ctx|\), every basic formula $bc \in Bc_\sigma$, and every $\delta_\cM \in l_{|\cM|(\sigma)}$, there exists a finitely presentable model (see Definition~\ref{def:fp object}) $\cM_{bc}$ such that

        \begin{enumerate}
            \item\label{enum:finiteness:1} for every morphism $\mu : \cN \to \cM$, $\cN \models_{|\mu|^*_\sigma(\delta_\cM)} \sigma.bc$ if and only if there exists a morphism \(\mu_{bc} : \cM_{bc} \to \cN\) such that \(\sem{\cM_{bc}}(\sigma.bc) = |\mu \circ \mu_{bc}|^*_\sigma(\delta_\cM)\).
            \item\label{enum:finiteness:2} for every morphism $\mu : \cM \to \cN$,  there exists $\delta_\cN \in l_{|\cN|(\sigma)}$ such that $\delta_\cM \preceq_{|\cM|(\sigma)} |\mu|^*_\sigma(\delta_\cN)$ and satisfying the following property:
            $\cN \models_{\delta_\cN} \sigma.bc$ if and only if there exists a morphism \(\mu_{bc} : \cM_{bc} \to \cN\) such that \(\sem{\cM_{bc}}(\sigma.bc) \preceq_{|\cM_{bc}|(\sigma)} |\mu_{bc}|^*_\sigma(\delta_\cN)\).
        \end{enumerate}

    \end{description}
\end{definition}

Definition~\ref{def:filteredsemsyst} calls for some comments.
\begin{itemize}
    \item To illustrate the projection condition, let us assume that \(\cC\) is \(\Set\) and \(Prop_\cC\) is the functor \(\Sub\). Then, each \(\iota_i\) is some element \(a_i\), and \(\iota_I\) is the tuple \((a_1, \ldots, a_n)\). For each \(i\in I\), \(|p_{I,i}|^*_\sigma(\iota_i)\) contains all tuples of the form \((x_1,\ldots x_{i-1},a_i,x_{i+1}, \ldots, x_{n})\). Taking the intersection over \(I\), we retrieve exactly \((a_1, \ldots, a_n)\). The projection condition generalizes this property to arbitrary semantical systems.

    \item The last condition will prove helpful in dealing with filtered products. Indeed, filtered products are defined componentwise and are closed only under finite intersections. Together with the condition of sup-generation, the finiteness condition essentially means that we consider objects as being generated by finite generators (in a loose sense). For example, in the case of FOL over presheaves (see Example~\ref{ex:finiteness FOL} below), the two conditions ensure that the generators are functors $G: \cB^{op} \to \Set$ such that for all $b \in |\cB|$, $G(b)$ is a finite set. The finiteness condition will enable the proof of \Los's theorem in the case of basic formulas. This condition is an adaptation of conditions given by R. Diaconescu for institutions~\cite{Dia08}.
\end{itemize}

\begin{proposition}
    In the semantical systems of Sections~\ref{fol-semanticalsystem},~\ref{hol-semanticalsystem}, and~\ref{modal-semanticalsystem} for FOL, HOL, and ML, filtered models exist if the elementary topos $\cC$ has filtered products and projections of model products are epimorphisms.
\end{proposition}

\begin{proof}
    We show the property for the semantical system for FOL given in Section~\ref{fol-semanticalsystem}. The proofs for the other semantical systems are substantially similar. 
    
    Let $I$ be a set and $F$ be a filter over $I$. Let $(\cM_i)_{i \in I}$ be a family of models in $Mod$. We define the filtered product $\prod_F \cM$ of $(\cM_i)_{i \in I}$ as follows:
    \begin{itemize}
        \item for every $s \in S$, $(\prod_F M)_s$ is the filtered product of $(M_{i_s})_{i \in I}$.
        \item for every function name $f : s_1 \times \ldots \times s_n \to s$, by the universal property of colimit, $f^{\prod_F \cM}$ is the unique morphism such that the following diagram:
        \[\begin{tikzcd}[column sep=3em]
            {\prod_I M_{i_{s_1}} \times \ldots \times \prod_I M_{i_{s_n}}} & \prod_I M_{i_{s_n}} \\
            {\prod_F M_{s_1} \times \ldots \times \prod_F M_{s_n}} & \prod_F M_s
            \arrow["{f^{\prod_I \cM_i}}", from=1-1, to=1-2]
            \arrow["{\mu_{I_s}}", from=1-2, to=2-2]
            \arrow["{(\mu_{I_{s_1}},\ldots,\mu_{I_{s_n}})}"', from=1-1, to=2-1]
            \arrow["{f^{\prod_F \cM}}"', from=2-1, to=2-2]
        \end{tikzcd}\]
            commutes.
        \item for every relation name $r:s_1 \times \ldots \times s_n$, $r^{\prod_F \cM}$ 
        is the subobject $O_F \mto \prod_F M_{s_1} \times \ldots \times \prod_F M_{s_n}$ where $O_F = \prod_F O$ is the filtered product of the family $(\dom(r^{\cM_i}))_{i \in I}$.

    \end{itemize}

    Hence, the family $\mu = (\mu_J)_{J \in F}$ forms a cocone $A_F \Rightarrow \prod_F \cM$ where $A_F : F \to Mod;J \mapsto \prod_J \cM_i,J \subseteq J' \mapsto p_{J',J}$. Let $\nu : A_F \Rightarrow \cN$ be another cocone. As $\cC$ has filtered products, there is a unique morphism $\theta : \prod_F M \to N$. Let us show that $\theta$ is a morphism in $Mod$. By definition, for every $J \in F$ and for all \(x:\prod_J M_{j_{s_1}} \times \ldots \times \prod_J M_{j_{s_n}}\), \(\theta\) satisfies  \(\theta(\mu_J(x)) = \nu_J(x)\).
    
    Let $f : s_1 \times \ldots \times s_n \to s$ be a function name. We want to show that $\theta \circ f^{\prod_F \cM} = f^{\cN} \circ \theta$, which essentially amounts to showing that the following diagram commutes.
    \[\begin{tikzcd}[column sep=5em, row sep=3em]
    	{\prod_I M_{i_{s_1}} \times \ldots \times \prod_I M_{i_{s_n}}} & {\prod_I M_{i_{s_n}}} \\
    	{\prod_F M_{s_1} \times \ldots \times \prod_F M_{s_n}} & {\prod_F M_s} \\
    	{N_{s_1} \times \ldots \times N_{s_n}} & {N_s}
    	\arrow["{f^{\prod_I \cM_i}}", from=1-1, to=1-2]
    	\arrow["{\mu_{I_s}}", from=1-2, to=2-2]
    	\arrow["{(\mu_{I_{s_1}},\ldots,\mu_{I_{s_n}})}", from=1-1, to=2-1]
    	\arrow["{(\theta_{I_{s_1}},\ldots,\theta_{I_{s_n}})}", from=2-1, to=3-1]
    	\arrow["{\theta_{I_s}}", from=2-2, to=3-2]
    	\arrow["{f^\cN}"', from=3-1, to=3-2]
    	\arrow["{f^{\prod_F \cM}}", from=2-1, to=2-2]
    	\arrow["{(\nu_{I_{s_1}},\ldots,\nu_{I_{s_n}})}"', curve={height=80pt}, from=1-1, to=3-1]
    	\arrow["{\nu_{I_s}}", curve={height=-30pt}, from=1-2, to=3-2]
    \end{tikzcd}\]
    Note that we already know that the top square commutes from the definition of $f^{\prod_F \cM}$ and we have the following equalities:
    \[\begin{array}{ll}
    \theta(f^{\prod_F \cM}(\mu_I(x))) & = \theta(\mu_I(f^{\prod_I \cM_i}(x))) \\
                                      & = \nu_I(f^{\prod_I \cM_i}(x)) \\
                                      & = f^{\cN}(\nu_I(x)) \\
                                      & = f^{\cN}(\theta(\mu_I(x)))
    \end{array}\]
    Hence, we have that $\theta \circ f^{\prod_F \cM} \circ \mu_I = f^{\cN} \circ \theta \circ \mu_I$. Let us show that $\mu_I$ is an epimorphism. So let us suppose $f,g : \prod_F \cM \to X$ such that $f \circ \mu_I = g \circ \mu_I$. Because for every $J \in F$, $p_{I,J}$ is an epimorphism, we deduce that $f \circ \mu_J = g \circ \mu_J$. Now, as $\prod_F \cM$ is a filtered product, $(\mu_J)_{J \in F}$ is a jointly epic family, and then $f = g$. From this, we can then conclude that $\theta \circ f^{\prod_F \cM} = f^{\cN} \circ \theta$.

    \medskip
    Let $r:s_1 \times \ldots \times s_n \in R$. Let $r^{\prod_I \cM_i} : O_I \mto \prod_I \cM_{i_{s_1}} \times \ldots \times \cM_{i_{s_n}}$, and $r^{\cN} : O_N \mto N_{s_1} \times \ldots \times N_{s_n}$. By the fact that $\nu$ is a morphism, we can write:
    \[
    O_N = \{\theta(\mu_I(x_I)) \mid x_I \in r^{\prod_I \cM_I}\}
    \]
    Hence, the following formula in the internal language of $\cC$ is satisfied:
    $$\forall x_I \in \prod_I \cM_i,\mu_I(x_i) \in r^{\prod_F \cM} \Rightarrow \theta(\mu_I(x_I)) \in r^{\cN}$$
    which proves that there exists a morphism $O_F \to O_N$ such that the diagram
    \[\begin{tikzcd}[column sep=4em]
        O_F &  {\prod_F M_{s_1} \times \ldots \times \prod_F M_{s_n}} \\
        O_N & {N_{s_1} \times \ldots \times N_{s_n}}
        \arrow["{r^{\prod_F \cM}}", tail, from=1-1, to=1-2]
        \arrow["{(\theta_{s_1},\ldots,\theta{s_n})}", from=1-2, to=2-2]
        \arrow[""', from=1-1, to=2-1]
        \arrow["{r^{\cN}}"', tail, from=2-1, to=2-2]
    \end{tikzcd}\]
    commutes.
\end{proof}

\subsubsection{Finiteness condition}

The finiteness condition is satisfied in many situations, such as in the context of FOL over presheaves.

\begin{example}[Finiteness condition in FOL/HOL over presheaves]
\label{ex:finiteness FOL}
We consider the semantical system of Section~\ref{fol-semanticalsystem} over the FOL signature $\Sigma = (S,F,R)$, with the additional condition that the elementary topos is a category of presheaves $\hat{\cB}$. From Example~\ref{ex:supgeneretion:presheaves}, $\hat{\cB}$ is sup-generated, meaning that the finiteness condition is well-defined.

Let $\cM$ be a $\Sigma$-model, \([\vec{x}]\) a context with $\vec{x} = (x_1:s_1,\ldots,x_n:s_n)$, and $[\vec{x}].r(\vec{t})$ an atomic formula.

For the sake of simplicity, we consider a sup-generator $G: \cB^{op} \to \Set$ where for all $b \in |\cB|$, $G(b)$ is either a singleton or the empty set.
Iterating the process (i.e., adding recursively all the needed constants) yields the complete construction where each $G(b)$ is a finite set.
Hence, let $G$ be a generator of $\Sub(M_{s_1} \times \ldots \times M_{s_n})$ such that for every $b \in |\cB|$, $G(b) = \{(a^b_1,\ldots,a^b_n)\}$ or $G(b) = \emptyset$. Let $\Sigma' = (S,F',R)$ be the FOL signature obtained from $\Sigma$ by adding to $F$ the constants $a^b_{s_i} : s_i$ for $b \in |\cB|$ when $G(b) \neq \emptyset$. Let us define the $\Sigma'$-model $\emptyset$ as follows:

\begin{itemize}
    \item for every $s \in S$, $\emptyset_s : \cB^{op} \to \Set$ is the presheaf that maps
    \begin{itemize}
        \item \(b \in | \cB|\) to \(\{t:s \mid t \text{ is a } \Sigma'\text{-ground term of sort } s\}\) and,
        \item morphisms in \(\cB\) to the identify on \(\{t:s \mid t \text{ is a } \Sigma'\text{-ground term of sort } s\}\);
    \end{itemize}

    \item for every $f : s_1 \times \ldots \times s_n \to s \in F$, $f^\emptyset: \emptyset_{s_1} \times \ldots \times \emptyset_{s_n} \Rightarrow \emptyset_s$ is the natural transformation which for every $b \in |\cB|$, we have the mapping $f^\emptyset_b : \emptyset_{s_1}(b) \times \ldots \times \emptyset_{s_n}(b) \to \emptyset_s(b);(t_1,\ldots,t_n) \mapsto f(t_1,\ldots,t_n)$;
    \item $r^\emptyset : \cB^{op} \to \Set$ is the presheaf such for every $b \in |\cB|$, $r^\emptyset(b)$ is the set
    \[
        \set{(a^b_{s_1},\ldots,a^b_{s_n})}
        \cup
        \set{(t_1,\ldots,t_n) \mid \forall j, 1 \leq j \leq n, t_j:\mbox{$\Sigma$-ground term},\sem{\cM}_{[]}(t_j)_b(\mathbb{1}) = a^b_j}
    \]
    if $G(b) \neq \emptyset$, and $r^\emptyset(b) = \emptyset$ otherwise (with \([]\) being the empty context);

    \item for all $r' \neq r \in R$, $r'^\emptyset = b \mapsto \emptyset$.
\end{itemize}

We define $\cM_{[\vec{x}].r(\vec{t})}$ as the $\Sigma$-model obtained from $\emptyset$ by forgetting the interpretation of constants $a^b_{s_i}$.

First, we show that $\cM_{[\vec{x}].r(\vec{t})}$ is finitely presentable.
Consider a morphism $\mu : \cM_{[\vec{x}].r(\vec{t})} \to \cN$ where the $(\nu_i)_{i \in I}$ is a colimit of a directed diagram $(f_{i,j})_{(i < j) \in (I,\leq)}$ of models in $Mod$:
\[\begin{tikzcd}
	& {\cN_i} && {\cN_j} \\
	{\cM_{[\vec{x}].r(\vec{t})}} && \cN
	\arrow["{f_{i,j}}", from=1-2, to=1-4]
	\arrow["{\nu_j}", from=1-4, to=2-3]
	\arrow["{\nu_i}"', from=1-2, to=2-3]
	\arrow["\mu"', from=2-1, to=2-3]
\end{tikzcd}\]
We consider the presheaf $H : \cB^{op} \to \Set;b \mapsto \mu_b(r^\emptyset(b))$ and the associated subobject $\iota : H \Rightarrow N_{s_1} \times \ldots \times N_{s_n}$. By construction, we have that $\cN \models_\iota [\vec{x}].r(\vec{t})$, and then for at least one $i \in I$ we have that $\cN_i \models_\delta [\vec{x}].r(\vec{t})$ where $\delta : K \Rightarrow \cN_{i_{s_1}} \times \ldots \times \cN_{i_{s_n}}$ is the subobject such that for every $b \in |\cB|$, $K(b) = \nu_{i_b}^{-1}(H(b))$. This gives the desired model morphism $\cM_{[\vec{x}].r(\vec{t})} \to \cN_i$. 

\medskip
\emph{Subcondition~\ref{enum:finiteness:1} of the finiteness condition.}
Let us consider a morphism $\mu : \cN \to \cM$ and define the morphism $\nu : \cM_{[\vec{x}].r(\vec{t})} \to \cN$ as the mapping $(\nu_s)_b : t \mapsto \sem{\cN}_{[\vec{x}]}(t)$, for every $s \in S$ and for every $b \in |\cB|$. The mapping is constructed with the convention that if $s \in \{s_1,\ldots,s_n\}$ and $t$ is $a^b_{s_i}$, then $\sem{\cN}_{[\vec{x}]}(t)$ is chosen in $\mu_b^{-1}(a^b_i)$. Then, the desired equivalence holds by construction.

\emph{Subcondition~\ref{enum:finiteness:2} of the finiteness condition.}
Let us consider a morphism $\mu : \cM \to \cN$ and define the presheaf $G' : \cB^{op} \to \Set$ which for every $b \in |\cB|$ associates $\{(\mu(a^b_1),\ldots,\mu(a^b_n))\}$ if $G(b) \neq \emptyset$, and $\emptyset$ otherwise. By construction, $G \subseteq |\mu|^*_{[\vec{x}]}(G')$. First, we suppose that $\cN \models_{G'} [\vec{x}].r(\vec{t})$. We define the morphism $\mu_{bc} : \cM_{[\vec{x}].r(\vec{t})} \to \cN$ as the mapping $(\mu_{bc_s})_b : t \mapsto \sem{\cN}_{[\vec{x}]}(t)$, for every $s \in S$ and for every $b \in |\cB|$. The mapping is constructed with the convention that if $s \in \{s_1,\ldots,s_n\}$ and $t$ is $a^b_{s_i}$, then $\sem{\cN}_{[\vec{x}]}(t) = \mu_b(a^b_i)$. The definition yields $\sem{\cM_{bc}}([\vec{x}]).r(\vec{t})) \subseteq |\mu \circ \mu_{bc}|^*_{[\vec{x}]}(G')$, i.e., the first direction of the subcondition~\ref{enum:finiteness:2}. Secondly, we suppose that there exists a morphism $\mu_{bc} : \cM_{bc} \to \cN$ such that $\sem{\cM_{bc}}([\vec{x}]).r(\vec{t}) \subseteq |\mu \circ \mu_{bc}|^*_{[\vec{x}]}(G')$. By definition of the morphism, this means that $G' \subseteq \sem{\cN}([\vec{x}]).r(\vec{t})$, and then $\cN \models_{G'} [\vec{x}].r(\vec{t})$.

\medskip
In HOL, the construction for atomic formulas of the form $t \in_A t'$ is identical. This construction could have also been extended to any semantical system for FOL/HOL over a locally finitely presentable topos. 
\end{example}

\begin{example}[Finiteness condition for ML over presheaves]
Here, we consider the semantical system of Section~\ref{modal-semanticalsystem} over a functor $F : \cC \to \cC$ and a category of presheaves $\widehat{\cB}$. Let $((X,\alpha),\nu)$ be a $F$-model (with $\nu$ being a mapping $PV \to \Sub(X)$ as in Example~\ref{ex:basic formula in modal logic}) and let $p \in PV$ be a propositional variable. Let $G : \cB^{op} \to \Set$ be a sup-generator in $l_X$. We consider the $F$-model $\cM_p = ((G,\alpha_G),\nu_G)$ where:
\begin{itemize}
    \item $\alpha_{G_b} : G(b) \to F(G)(b);x \mapsto \alpha_b(x)$
    \item $\nu_G : PV \to \Sub(X) ; p' \mapsto
    \left\{\begin{array}{ll}
        G                       & \text{ if } p' = p\\
        (b \mapsto \emptyset)   & \text{ if } p' \neq p\\
    \end{array}\right.
    $
\end{itemize}
The two equivalences in the conditions of the finiteness definition are easy to prove. 
\end{example}

\begin{example}[Finiteness condition in institutions]
We consider the semantical system $\cS_{\cI}(\Sigma)$ of Section~\ref{subsubsec:semantical system for institutions} for a signature $\Sigma \in |Sig|$ of an institution $\cI$ satisfying all the expected hypotheses (i.e., existence of small products in $Mod(\Sigma)$, and the forgetful functor $Mod(\chi)$ for any quasi-representable signature morphism $\chi$ creates small products). %
Here, we consider that the basic sentences are finitary, similar to~\cite{Dia08}. This means that there exists a finitely presentable $\Sigma$-model $\cM_{bc}$ such that:
\[\cM \models_\Sigma bc \; \mbox{iff} \; \mbox{there exists $\mu_{bc} : \cM_{bc} \to \cM$}\]
Let $\cM \in |Mod|$ be a model, $bc \in Bc_\chi$ a finitary basic formula with $\chi : \Sigma \to \Sigma'$, and $\cM' \in |Mod(\Sigma')|$ such that $Mod(\chi)(\cM') = \cM$.\footnote{Sup-generators for $Prop(|\cM|(\chi))$ are singletons.} In the following, we assume that $Mod(\chi)$ preserves finitely presentable models. As $bc$ is finitary, there exists a finitely presentable model $\cM'_{bc} \in |Mod(\Sigma')|$ such that $\cM' \models_{\Sigma'} bc$. \\ Let us set $\cM_{bc} = Mod(\chi)(\cM'_{bc})$.

\medskip
\emph{Subcondition~\ref{enum:finiteness:1} of the finiteness condition.}
Let us consider a morphism $\mu : \cN \to \cM$, and suppose that for every $\cN' \in (\mu,Id_{\Sigma'})^{-1}(\{\cM'\})$, $\cN' \models_{\Sigma'} bc$.
As $bc$ is basic (within the meaning of institutons~\cite{Dia08}), there is a morphism $\cM'_{bc} \to \cN'$, which yields a morphism $\mu_{bc} : \cM_{bc} \to \cN$.
Then, it follows from the definition of basic formulas in institutions (in the sense of Example~\ref{ex:basic formulas in Institutions}) that:
\[
\sem{\cM_{bc}}(\chi.bc) = \{\cM''_{bc} \in |Mod(\Sigma')| \mid \cM''_{bc} \models_{\Sigma'} bc \; \mbox{and} \; Mod(\chi)(\cM''_{bc}) = \cM_{bc}\}
\]
Hence, $\cM'_{bc} \to \cN'$ is the same unique extension of 
\[
\mu_{bc} : Mod(\chi)(\cM'_{bc}) = Mod(\chi)(\cM''_{bc}) \to \cN
\]
which leads to $\cM'_{bc} = \cM''_{bc}$.
We obtain that $\sem{\cM_{bc}}(\chi.bc) = \{\cM'_{bc}\}$. By the same arguments, we can also show that $|\mu \circ \cM_{bc} \to \cN|^*_\chi(\{\cM'\}) = \{\cM'_{bc}\}$.\\
For the opposite implication, if we suppose that for every $\cN' \in (\mu,Id_{\Sigma'})^{-1}(\{\cM'\})$, we have a morphism $\cM'_{bc} \to \cN'$, then as $bc$ is basic, we have that $\cN' \models _{\Sigma'} bc$, and then $\cN \models_{\{\cM'\}} bc$.

\medskip
\emph{Subcondition~\ref{enum:finiteness:2} of the finiteness condition.}
Let us consider a morphism $\mu: \cM \to \cN$. Let $\mu' : \cM' \to \cN'$ be the unique $\chi$-extension of $\mu$, and set $\delta = \{\cN'\}$. By construction, it holds that $\cM' \in (\mu,Id_{\Sigma'})^{-1}(\delta)$.  If $\cN' \models_{\Sigma'} bc$, then there is a morphism $\mu'_{bc} : \cM'_{bc} \to \cN'$, and then a morphism $\cM_{bc} \to \cN$. For the same reason as previously, we have that $\sem{\cM_{bc}}(\chi.bc) = \{\cM'_{bc}\} = |\mu_{bc}|^*_\chi(\delta)$. \\
Similarly, if we suppose that there is a morphism $\cM_{bc} \to \cN$, then it yields a morphism $\cM'_{bc} \to \cN'$.
We can deduce that $\cN' \models_{\Sigma'} bc$ and conclude that $\cN \models_\delta bc$.
\end{example}

\subsection{Filterable quantifiers}
\label{subsec:los:quantifiers}

We required a filtered semantical system to have the covering property, i.e., that \(\cC\) is covered. This covering property ensures that we can properly interpret disjunctive formulas. It naturally translates into a similar condition on the quantifiers.

\begin{definition}[Filterable quantifier]
\label{def:filterable quantifier}
Let $\mu : \cM \to \cM'$ be a model morphism. Let $\Qf$ be a n-ary quantifier where $f : \sigma \to \tau$ is a context morphism.
$\cQ f$ is {\bf filterable} when for every $\delta \in l_{|\cM|(\tau)}$ and $\iota_1,\ldots,\iota_n \in Prop_\cC(|\cM|(\sigma))$, if $\delta \preceq_{|\cM|(\tau)} \cQ f_\cM(\iota_1,\ldots,\iota_n)$, then  there exist $\delta_1,\ldots,\delta_n \in l_{|\cM|(\sigma)}$ such that for every $j$, $1 \leq j \leq n$,
\begin{itemize}
	\item $\delta_j \preceq_{|\cM|(\sigma)} \iota_j$, and
	\item $\delta = \Qf_{\cM}(\delta_1,\ldots,\delta_n)$.
\end{itemize}
\end{definition}

\begin{proposition}
\label{prop:FOL and HOL existential quantifiers are directly filterable}
Let $\mathcal{C}$ be a sup-generated topos. FOL and HOL existential quantifiers are filterable if generators are preserved by isomorphisms.
\end{proposition}

\begin{proof}
Let \(\gamma : [\vec{x}] \to [\vec{y}]\) be a morphism in \(\Ctx\), \(\delta_y \in l_{|\cM|([\vec{y}])}\) and \(\iota_x \in Sub(|\cM|([\vec{x}]))\) such that \(\delta_y \preceq_{|\cM|([\vec{y}])} \exists \gamma_\cM(\iota_x)\).
For a variable \(v_x:|\cM|([\vec{x}])\), we consider
\[
[v_x] = \set{v:|\cM|([\vec{x}]) \mid \sem{\cM}_{[\vec{x}]}(\gamma)(v) = \sem{\cM}_{[\vec{x}]}(\gamma)(v_x)}
\]
using the definition of \(\sem{\cM}_{[\vec{x}]}(\gamma)\) from Section~\ref{subsec:internallogic}.
Now, we consider \(\delta_x = \set{[v_x] \mid v_x \in \iota_x ~\text{and}~  \sem{\cM}_{[\vec{x}]}(\gamma)(v_x) \in \delta_y}\). By construction, \mbox{\(\delta_x \preceq_{|\cM|([\vec{x}])} \iota_x\),} \(\delta_y = \exists \gamma_\cM(\delta_x)\), and \(\delta_x\) is a generator.
Indeed, it holds that for all \(v_y:|\cM|([\vec{y}])\), if \(v_y \in \delta_y\), then there exists a unique \([v_x] \in \delta_x\) such that for all \(v \in [v_x]\), \(\sem{\cM}(\gamma)(v) = y\). The existence is given by the fact that \(\delta_y \preceq_{|\cM|([\vec{y}])} \exists \gamma_\cM(\iota_x)\) and the uniqueness by the construction of \([v_x]\). Therefore \(\delta_x \simeq \delta_y\), as an isomorphism of their domain in \(\cC\).

\end{proof}

\begin{proposition}
Institution existential quantifiers are filterable.
\end{proposition}

\begin{proof}
Let $(\cM_i)_{i \in I}$ be a family of models. Let $\theta : \chi_2 \to \chi_1$ be a context morphism. 
Let us suppose that $S \subseteq \exists \theta_{\cM}(S')$ with $S = \{\cM_1\} \subseteq |\cM|(\chi_1)$ and $S' \subseteq |\cM|(\chi_2)$. Let us define the set $T \subseteq |\cM|(\chi_2)$ as follows:
$$T = \{\cM_2\}$$
where $\cM_2$ is any $\Sigma_2$-model of $S'$ such that $Mod(\theta)(\cM_2) = \cM_1$. 
By construction, we directly have that $T \subseteq S'$. Likewise,  the fact that $S = \exists \theta_\cM(T)$ is obvious by hypothesis.
\end{proof}

\medskip

\subsection{Filterable pullback functors}
\label{subsec:los:pullbacks}

\begin{definition}[Filterable pullback functors]

$f$ is said {\bf filterable} when for all $\delta_\sigma \in l_{|\cM|(\sigma)}$ and all $\iota \in Prop_\cC(|\cM|(\tau))$, if $\delta_\sigma \preceq_{|\cM|(\sigma)} |\cM|(f)^*(\iota)$, then there exists $\delta_\tau \in l_{|\cM|(\tau)}$ such that
\begin{itemize}
    \item $\delta_\tau \preceq_{|\cM|(\tau)} \iota$
    \item $\delta_\sigma = |\cM|^*(f)(\delta_\tau)$
\end{itemize}
\end{definition}

\begin{proposition}
Let $\cC$ be a sup-generated topos. FOL and HOL context morphisms are filterable for any model morphism. 
\end{proposition}

\begin{proof}
The proof is similar to the proofs of Propositions~\ref{prop:FOL and HOL existential quantifiers are directly filterable}, replacing
\(\exists \gamma_\cM\)
by
\(|\cM|(\gamma)\).

\end{proof}

\subsection{\Los's theorem for abstract categorical logics}
\label{subsec:los:theorem}

For comparison, we recall the standard (set-theoretic) version of \Los's theorem.

\begin{theorem}[\Los's Theorem~\cite{los_quelques_1955}]
\label{th:Los:set-theory}
Let \((\cM_i)_{i\in I}\) be an \(I\)-indexed family of nonempty \(\Sigma\)-structures, and let \(F\) be an ultrafilter on \(I\). Let \(\prod_F \cM\) be the ultraproduct of \((\cM_i)_{i\in I}\) with respect to \(F\). Since each \(\cM_i\) is nonempty, the ultraproduct \(\prod_F \cM\) is the quotient of \(\prod_{i \in I} \cM_i\) by the equivalence relation identifying \(I\)-sequences that coincide on a set of indices belonging to \(F\). Let \((a^k_i)_{i \in I}\) be \(I\)-sequences for \(k \in \set{1, \ldots, n}\), with \([a^k]\) denoting their equivalence class. Then for each \(\Sigma\)-formula \(\varphi\),
\[
    \prod_F \cM \models \varphi([a^1], \ldots [a^n]) ~\mbox{iff}~ \set{j \in I \mid \cM_j \models \varphi(a^1_j, \ldots a^n_j)} \in F.
\]
\end{theorem}

We now state and prove our main result: an extension of \Los's theorem for abstract categorical logics. In our abstract framework, the right part of the equivalence is essentially the same, i.e., the set of indices $i$ such that the formula holds in $\cM_i$ is an element of the ultrafilter $F$. However, considering the set of equivalence classes can no longer be achieved pointwise, leading to a reformulation of the left part of the equivalence.

\begin{theorem}[Abstract \Los's Theorem]
\label{th:Los}
Let \(\cL = (\cS,\cQ,Bc)\) be a logic such that 
\begin{itemize}
	\item $\cS = (\cC,\Ctx,Mod,|\_|)$ is a filtered semantical system, and
	\item for all ultrafilters $F$ over a set  $I$, and all families of models $(\cM_i)_{i \in I}$, quantifiers and pullback functors are filterable and distributing over morphisms $\mu_J$ and $p_{K,J}$ such that $J \subseteq K$, for all $K,J \in F$.
\end{itemize}

Let $F$ be an ultrafilter over a set I. Let $(\cM_i)_{i \in I}$ be a family of models. For all formulas \(\sigma.\varphi \in \cF_\cL\), and all $\delta_I \in l_{|\prod_I \cM_i|(\sigma)}$, we have:
\[
\delta_I \preceq_{|\prod_I \cM_i|(\sigma)} |\mu_I|^*_\sigma(\sem{\prod_F \cM}(\sigma.\varphi)) \; \mbox{iff} \; \bigset{i \in I \mid \delta_I \preceq_{|\prod_I \cM_i|(\sigma)} |p_{I,i}|^*_\sigma(\sem{\cM_i}(\sigma.\varphi))} \in F
\]
\end{theorem}

\begin{proof}
    The proof is done by structural induction on $\varphi$. We will not subscript the order of the Heyting algebra by the object of the prop-category to simplify the expressions; it can be deduced from the context.
    \begin{itemize}
        \item The case of $\top$ is obvious, and the case of $\bot$ is a consequence of the fact that $\emptyset \notin F$, and generators cannot be the lowest bound of Heyting algebra.
        \item $\varphi$ is $bc \in Bc_\sigma$.

        \medskip
        ($\Rightarrow$) Suppose that $\delta_I \preceq |\mu_I|^*_\sigma(\sem{\prod_F \cM}(\sigma.bc))$.
        By the condition of Definition~\ref{def:is supgenerated}, there exists $\delta_F \in l_{|\prod_F \cM|(\sigma)}$ such that $\delta_I \preceq |\mu_I|^*_\sigma(\delta_F)$ and $\delta_F \preceq \sem{\prod_F \cM}(\sigma.bc)$.
        Since \(\prod_F \cM \models_{\delta_F} \sigma.bc \), subcondition~\ref{enum:finiteness:1} of the finiteness condition (see Definition~\ref{def:filteredsemsyst}) with \(Id_{ \prod_F \cM}\) ensures that there exists a finitely presentable model $\cM_{bc} \in |Mod|$ and a morphism $\mu_{bc} : \cM_{bc} \to \prod_F \cM$ such that $\sem{\cM_{bc}}(\sigma.bc) = |\mu_{bc}|^*_\sigma(\delta_F)$. As $\cM_{bc}$ is finitely presentable (see Definition~\ref{def:fp object}), there exists a non-empty set \(J\in F\) and a morphism $\mu : \cM_{bc} \to \prod_J \cM_j$ such that \(\mu_{bc} = \mu_J \circ \mu\).
        Hence, we have that $\sem{\cM_{bc}}(\sigma.bc) = |\mu_J \circ \mu|^*_\sigma(\delta_F)$.
        From the subcondition~\ref{enum:finiteness:1} (reverse implication) of the finiteness condition (see Definition~\ref{def:filteredsemsyst}) with \(\mu_J\), it follows that \(|\mu_J|^*_\sigma(\delta_F) \preceq \sem{\prod_{j\in J} \cM_j}(\sigma.bc)\).
        By the condition of Definition~\ref{def:basic formulas} on basic formulas, we obtain that, for all \(j\) in \(J\), \(|\mu_J|^*_\sigma(\delta_F) \preceq |p_{J,j}|^*_\sigma(\sem{\cM_j}(\sigma.bc))\).
        Since \(|p_{I,J}|^*_\sigma\) is a morphism of Heything algebras, for all \(j\) in \(J\), \(|p_{I,J}|^*_\sigma \circ |\mu_J|^*_\sigma(\delta_F) \preceq |p_{I,j}|^*_\sigma(\sem{\cM_j}(\sigma.bc))\).
        Now, because \(\mu_I = \mu_J \circ p_{I,J}\), we have that \(|\mu_I|^*_\sigma = |p_{I,J}|^*_\sigma \circ |\mu_J|^*_\sigma\) and, therefore, \(\delta_I \preceq |p_{I,J}|^*_\sigma \circ |\mu_J|^*_\sigma (\delta_F)\).
        Thus, for all for all \(j\) in \(J\), \(\delta_I \preceq |p_{I,j}|^*_\sigma(\sem{\cM_j}(\sigma.bc))\). In particular, since \(F\) is a filter, the set
        \(
            \bigset{i \in I \mid \delta_I \preceq_{|\prod_I \cM_i|(\sigma)} |p_{I,i}|^*_\sigma(\sem{\cM_i}(\sigma.bc))}
        \) containing \(J\) is in \(F\).

        \medskip
        ($\Leftarrow$) Suppose that $J = \{i \in I \mid \delta_I \preceq |p_{I,i}|^*_\sigma(\sem{\cM_i}(\sigma.bc))\} \in F$. Then \(J\) is not empty and for each \(j\in J\), \(\delta_I \preceq |p_{I,j}|^*_\sigma(\sem{\cM_j}(\sigma.bc))\).
        By the condition of Definition~\ref{def:is supgenerated} (since \(\cC\) is sup-generated), for every $j \in J$, there exists $\delta_j \in l_{|\cM_j|(\sigma)}$  such that $\delta_I \preceq |p_{I,j}|^*_\sigma(\delta_j)$ and \(\delta_j \preceq \sem{\cM_j}(\sigma.bc)\).
        By the projection condition of Definition~\ref{def:filteredsemsyst} on the \(J\)-indexed family \((\delta_j)_{j \in J}\), there exists \(\iota_J \in Prop_\cC(|\prod_J \cM_j|(\sigma))\) such that \(\downarrow\!\iota_J = \bigcap_{j \in J} \downarrow\!|p_{J,j}|^*_\sigma(\delta_j)\).
        From Definition~\ref{def:supgenerator}, it follows that \(\iota_J = \bigvee\!\downarrow\!\iota_J\) and then for all \(j\) in \(J\), \(\iota_J \preceq |p_{J,j}|^*_\sigma(\delta_j) \preceq |p_{J,j}|^*_\sigma(\sem{\cM_j}(\sigma.bc))\).
        By the condition of interpretability in Definition~\ref{def:basic formulas}, we then have that $\iota_J \preceq \sem{\prod_J \cM_j}(\sigma.bc)$. 
        Additionally, for all \(j \in J\), \(\delta_I \preceq |p_{I,j}|^*_\sigma(\delta_j)\), meaning that \(\delta_I \in \set{\delta \in l_{|\prod_I \cM_i|(\sigma)} \mid \forall j \in J,~ \delta \preceq |p_{I,J}|^*_\sigma \circ |p_{J,j}|^*_\sigma (\delta_j)}\).
        Since \(|p_{I,J}|^*_\sigma\) is a morphism of Heything algebras and \(\iota_J = \bigvee \set{\delta \in l_{|\prod_J \cM_j|(\sigma)} \mid \forall j \in J,~ \delta \preceq |p_{J,j}|^*_\sigma (\delta_j)}\), it follows that \(\delta_I \preceq |p_{I,J}|^*_\sigma(\iota_J)\)
        Therefore, by the condition of Definition~\ref{def:is supgenerated}, there exists $\delta_J \in l_{|\prod_J \cM_j|(\sigma)}$ such that $\delta_I \preceq |p_{I,J}|^*_\sigma(\delta_J)$ and $\delta_J \preceq \iota_J$. In particular, \(\prod_J \cM_j \models_{\delta_J} \sigma.bc\).
        By the subcondition~\ref{enum:finiteness:1} of the finiteness condition (see Definition~\ref{def:filteredsemsyst}) with \(Id_{ \prod_J \cM_j}\), there exists a finitely presentable model $\cM_{bc} \in |Mod|$ and a morphism $\mu_{bc} : \cM_{bc} \to \prod_J \cM_j$ such that $\sem{\cM_{bc}}(\sigma.bc) = |\mu_{bc}|^*_\sigma(\delta_J)$.
        Now, \(\mu_J\) is a morphism \(\prod_J \cM_j \to \prod_F \cM\), and the subcondition~\ref{enum:finiteness:2} of the finiteness condition yields a generator $\delta_F \in l_{|\prod_F \cM|(\sigma)}$ such that $\delta_J \preceq |\mu_J|^*_\sigma(\delta_F)$.
        The morphism \(\mu_J \circ \mu_{bc} : \cM_{bc} \to \prod_F \cM\) ensures that \(\sem{\cM_{bc}}(\sigma.bc) = |\mu_{bc}|^*_\sigma(\delta_J) \preceq |\mu_J \circ \mu_{bc}|^*_\sigma(\delta_F)\), and then \(\prod_F \cM \models_{\delta_F} \sigma.bc\).
        From \(\delta_F \preceq \sem{\prod_F \cM}(\sigma.bc)\), \(\delta_J \preceq |\mu_J|^*_\sigma(\delta_F)\), and \(\delta_I \preceq |p_{I,J}|^*_\sigma(\delta_J)\), we conclude that $\delta_I \preceq |\mu_I|^*_\sigma(\sem{\prod_F \cM}(\sigma.bc))$.

        \item The case of conjunctions is obvious. 

        \item $\varphi$ is $\psi \vee \chi$.

        \medskip
        ($\Rightarrow$) Suppose that $\delta_I \preceq |\mu_I|^*_\sigma(\sem{\prod_F \cM}(\sigma.\psi \vee \chi))$. By the covering property, this means that there exists $\delta^\psi_I,\delta^\chi_I \in l_{|\prod_I \cM_i|(\sigma)}$ such that $\delta_I = \delta^\psi_I \vee \delta^\chi_I$, $\delta^\psi_I \preceq |\mu_I|^*_\sigma(\sem{\prod_F \cM}(\sigma.\psi))$, and $\delta_I \preceq |\mu_I|^*_\sigma(\sem{\prod_F \cM}(\sigma.\chi))$. By the induction hypothesis, we have that
        \begin{itemize}
            \item $J = \{i \in I \mid \delta^\psi_I \preceq |p_{I,i}|^*_\sigma(\sem{\cM_i}(\sigma.\psi))\} \in F$
            \item $K = \{i \in I \mid \delta^\chi_I \preceq |p_{I,i}|^*_\sigma(\sem{\cM_i}(\sigma.\chi))\} \in F$
        \end{itemize}

        Let us set $L = J \cup K$. Then, we have for every $l \in L$ that $\delta_I \preceq |p_{I,l}|^*_\sigma(\sem{\cM_l}(\sigma.\psi \vee \chi)$. 

        \medskip
        ($\Leftarrow$) Suppose that $J = \{i \in I \mid \delta_I \preceq |p_{I,i}|^*_\sigma(\sem{\cM_i}(\sigma.\psi \vee \chi))\} \in F$. By the projection condition, there exists $\delta^\psi_J,\delta^\chi_J \in Prop_\cC(|\prod_J \cM_j|(\sigma))$ such that $\downarrow\!\delta^\psi_J = \bigcap_{j \in J} \downarrow\!|p_{J,j}|^*_\sigma(\sem{\cM_j}(\sigma.\psi))$ and $\downarrow\!\delta^\chi_J =  \bigcap_{j \in J} \downarrow\!|p_{J,j}|^*_\sigma(\sem{\cM_j}(\sigma.\chi))$, from which we can deduce that $\delta_I \preceq |p_{I,J}|^*_\sigma(\delta^\psi_J) \vee |p_{I,J}|^*_\sigma(\delta^\chi_J)$.
        By the covering property, there exists $\delta^\psi_I,\delta^\chi_I \in l_{\prod_I \cM_i|(\sigma)}$ such that $\delta^\psi_I \preceq |p_{I,J}|^*_\sigma(\delta^\psi_J)$, $\delta^\chi_I \preceq |p_{I,J}|^*_\sigma(\delta^\chi_J)$, and \(\delta_I = \delta^\chi_I \vee \delta^\psi_I\).
        By construction, for all \(j\) in \(J\), \(\delta_I^\chi \preceq |p_{I,j}|^*_\sigma(\sem M_j (\sigma.\chi)\) and \(\delta_I^\psi \preceq |p_{I,j}|^*_\sigma(\sem M_j (\sigma.\psi)\), meaning that the sets \(\bigset{i \in I \mid \delta^\psi_I \preceq |p_{I,i}|^*_\sigma(\sem{\cM_i}(\sigma.\psi))}\) and \(\bigset{i \in I \mid \delta^\chi_I \preceq |p_{I,i}|^*_\sigma(\sem{\cM_i}(\sigma.\chi))}\) are in \(F\).
        By the induction hypothesis, we then have that $\delta^\psi_I \preceq |\mu_I|^*_\sigma(\sem{\prod_F \cM}(\sigma.\psi))$ and $\delta^\chi_I \preceq |\mu_I|^*_\sigma(\sem{\prod_F \cM}(\sigma.\chi))$, and then $\delta_I \preceq |\mu_I|^*_\sigma(\sem{\prod_F \cM}(\sigma.\psi \vee \chi))$.

        \item $\varphi$ is $\psi \Rightarrow \chi$.

        \medskip
        ($\Rightarrow$) Suppose that $\{i \in I \mid \delta_I \preceq |p_{I,i}|^*_\sigma(\sem{\cM_i}(\sigma.\psi \Rightarrow \chi))\} \notin F$. This means that the set $J = \{i \in I \mid \delta_I \preceq |p_{I,i}|^*_\sigma(\sigma.\psi)),\delta_I \not\preceq |p_{I,i}|^*_\sigma(\sigma.\chi))\} \in F$. By the induction hypothesis, we then have that $\delta_I \preceq |\mu_I|^*_\sigma(\sem{\prod_F \cM}(\sigma.\psi))$. By contradiction, let us suppose that $\delta_I \preceq |\mu_I|^*_\sigma(\sem{\prod_F \cM}(\sigma.\chi))$. By the induction hypothesis, this means that the set $K = \{i \in I \mid \delta_I \preceq |p_{I,i}|^*_\sigma(\sigma.\chi))\} \in F$. Let us set $L = J \cap K$. Then, we have that $L \subseteq \{i \in I \mid \delta_I \preceq |p_{I,i}|^*_\sigma(\sem{\cM_i}(\sigma.\psi \Rightarrow \chi))\}$ which is a contradiction, and then $\delta_I \not\preceq |\mu_I|^*_\sigma(\sem{\prod_F \cM}(\sigma.\psi \Rightarrow \chi))$.

        \medskip
        ($\Leftarrow$) Suppose that $\delta_I \not\preceq |\mu_I|^*_\sigma(\sem{\prod_F \cM}(\sigma.\psi \Rightarrow \chi))$. Then, $\delta_I \not\preceq |\mu_I|^*_\sigma(\sem{\prod_F \cM}(\sigma.\chi))$ and, by the induction hypothesis, it follows that $\{i \in I \mid \delta_I \preceq |p_{I,i}|^*_\sigma(\sem{\cM_i}(\sigma.\chi))\} \notin F$. Thus, we obtain that $\{i \in I \mid \delta_I \preceq |p_{I,i}|^*_\sigma(\sem{\cM_i}(\sigma.\psi \Rightarrow \chi))\} \notin F$. 

        \item $\varphi$ is $\cQ f(\varphi_1,\ldots,\varphi_n)$. 

        \medskip
        ($\Rightarrow$) Suppose that $\delta_I \preceq |\mu_I|^*_\tau(\sem{\prod_F \cM}(\tau.\cQ f(\varphi_1,\ldots,\varphi_n))$. As $\cQ f$ is distributing over $\mu_I$, we have that 
        \[
        \delta_I \preceq \cQ f_{\prod_I \cM_i}(|\mu_I|^*_\sigma(\sem{\prod_F \cM}(\sigma.\varphi_1)),\ldots,|\mu_I|^*_\sigma(\sem{\prod_F \cM}(\varphi_n)))
        \]
        Because $\cQ f$ is directly filterable, there exists generators $\delta^1_I,\ldots,\delta^n_I$ in $l_{|\prod_I \cM_i|(\sigma)}$ such that $\delta_I = \cQ f_{\prod_I \cM_i}(\delta^1_I,\ldots,\delta^n_I)$ and for every $j$, $1 \leq j \leq n$, $\delta^j_I \preceq |\mu_I|^*_\sigma(\sem{\prod_F \cM}(\sigma.\varphi_j))$. By the induction hypothesis, for every $j$, $1 \leq j \leq n$, the set $J_j = \{i \in I \mid \delta^j_I \preceq |p_{I,i}|^*_\sigma(\sem{\cM_i}(\sigma.\varphi_j))\}$ is in $F$. Let $L$ be $\bigcap_{j} J_j$. Then \(L\) is in \(F\) and, for every $l \in L$, for every $j$, $1 \leq j \leq n$, it holds that $\delta^j_I \preceq |p_{I,l}|^*_\sigma(\sem{\cM_l}(\sigma.\varphi_j))$. We can conclude that for all \(l \in L\), $\delta_I \preceq |p_{I,l}|^*_\tau(\sem{\cM_l}(\tau.\cQ f(\varphi_1,\ldots,\varphi_n))$. 

        \medskip
        ($\Leftarrow$) Suppose that $J = \bigset{i \in I \mid \delta_I \preceq |p_{I,i}|^*_\sigma(\sem{\cM_i}(\tau.\cQ f(\varphi_1,\ldots,\varphi_n)))} \in F$. By the projection condition, for every $k$, $1 \leq i \leq k$, for the family $\big(|p_{I,i}|^*_\sigma(\sem{\cM_i}(\sigma.\varphi_k))\big)_{j \in J}$, there exists $\delta^k_J \in Prop_\cC(|\prod_J \cM_j|(\sigma))$ such that $\downarrow\!\delta^k_J = \bigcap_{j \in J} \downarrow\!|p_{J,j}|^*_\sigma(\sem{\cM_j}(\sigma.\varphi_k))$, and then $\delta^k_J \preceq |p_{J,j}|^*_\sigma(\sem{\cM_j}(\sigma.\varphi_k))$, from which we have that $|p_{I,J}|^*_\sigma(\delta^k_J) \preceq |p_{I,j}|^*_\sigma(\sem{\cM_j}(\sigma.\varphi_k))$. As $\cQ f$ is filterable, there exists a generator \(\delta^k_I\) for every $k$, $1 \leq k \leq n$, such that $\delta^k_I \preceq |p_{I,J}|^*_\sigma(\delta^k_J)$ and $\delta_I = \cQ f_{\prod_I \cM_i}(\delta^1_I,\ldots,\delta^n_I)$. By the induction hypothesis, we have for every $k$, $1 \leq k \leq n$, that $\delta^k_I \preceq |\mu_I|^*_\sigma(\sem{\prod_F \cM}(\sigma.\varphi_k))$. As quantifiers are distributing over $\mu_I$, we can conclude that $\delta_I \preceq |\mu_I|^*(\sem{\prod_F \cM}(\tau.\cQ f(\varphi_1,\ldots,\varphi_n)))$.

        \item $\varphi$ is $f(\psi)$. The proof is similar to the proof for quantifiers.
    \end{itemize} 

\end{proof}

\subsection{Dual quantifiers}
\label{subsec:los:duality}

FOL universal quantifiers are not filterable following Definition~\ref{def:filterable quantifier}. For instance,  for FOL over $\Set$, given a generator defined by a singleton $\{[(a^{[\vec{y}]}_i)_{i \in I}]_{\equiv_F}\} \in l_{|\prod_F \cM|([\vec{y}])}$ such that\footnotemark{} $\prod_F \cM \models_{[(a^{[\vec{y}]}_i)_{i \in I}]} \forall \pi.\varphi$ with $\pi : [\vec{x}] \to [\vec{y}]$ a projection morphism in $\Ctx$ (i.e., with \(\vec{x} = \vec{y}.\vec{z}\) for some \(\vec{z}\)), we can define $\delta = \{[(a^{[\vec{y}]}_i,a^{[\vec{z}]}_i)_{i \in I}] \mid (a^{[\vec{z}]}_i)_{i \in I} \in |\prod_I \cM_i|([\vec{z}])\}$. This satisfies that $\delta \preceq_{|\prod_F \cM|([\vec{x}])} \sem{\prod_F \cM}([\vec{x}].\varphi)$ and $(a^{[\vec{y}]}_i)_{i \in I} = \forall \pi (\delta)$. The problem is that $\delta$ is not a generator, i.e., $\delta \notin l_{|\prod_F \cM|([\vec{x}])}$. Hence, to apply the induction hypothesis, we have to consider all the generators $\delta' \in l_{|\prod_F \cM|([\vec{x}])}$ such that $\delta' \preceq \delta$. By applying the induction hypothesis on $\delta'$, we obtain a set $K_{\delta'} \in F$. The problem is that the set $\{\delta' \in l_{|\prod_F \cM|} \mid \delta' \preceq \delta\}$ is very likely to be infinite, and filters are not closed under infinite intersections. Now, we know that $[\vec{y}].\exists \pi \neg \varphi \Rightarrow \neg \forall \pi \varphi$. Let us extend this implication to non-filterable quantifiers.

\footnotetext{$[(a_i)_{i \in I}]_{\equiv_F}$ is the equivalence class of all sequences of values $(b_i)_{i\in I}$ equivalent to $(a_i)_{i \in I}$ for the equivalence relation $\simeq_F$ associated to the filter $F$ on $I$.}

\medskip
In the following, we will say that a quantifier is {\bf globally filterable}  when it is filterable following Definition~\ref{def:filterable quantifier}, but the subobjects $\delta_i$ are not necessarily generators.  

\begin{definition}[Dual quantifier]
\label{def:dual quantifier}
Let $\Qf$ be a globally filterable quantifier with $f : \sigma \to \tau \in \Ctx$. $\Qf$ is said {\bf dual} if there exists a quantifier $\overline{Qf}$ satisfying the duality condition: for every model $\cM$, and every $\iota \in l_{|\cM|(\tau)}$, $\iota \not\preceq_{|\cM|(\tau)} \sem{\cM}(\tau.\Qf(\varphi_1,\ldots,\varphi_n))$ iff $\iota \preceq_{|\cM|(\tau)} \sem{\cM}(\tau.\overline{Qf}(\overline{\varphi}_1,\ldots,\overline{\varphi}_n))$  where there exists a subset $S \subseteq \{1,\ldots,n\}$ such that for every $j$, $1 \leq j \leq n$, if $j \in S$ then $\overline{\varphi}_j = \neg \varphi_j$, else $\overline{\varphi}_j = \varphi_j$.  
\end{definition}

We can extend \Los's theorem to dual quantifiers. Indeed, we have:

\begin{itemize}
    \item $(\Rightarrow)$ Let us suppose that $\{i \in I \mid \delta_I \preceq |p_{I,i}|^*_\tau(\sem{\cM_i}(\tau.\cQ f(\varphi_1,\ldots,\varphi_n))\} \notin F$. As $F$ is an ultrafilter, this means that $$\{i \in I \mid \delta_I \not\preceq |p_{I,i}|^*_\tau(\sem{\cM_i}(\tau.\cQ f(\varphi_1,\ldots,\varphi_n))\} \in F$$ 
    By duality, this means that there is \(S \subseteq \{1,\ldots,n\}\) such that
    $$\delta_I \preceq |p_{I,i}|^*_\tau(\sem{\cM_i}(\tau.\overline{\Qf}(\overline{\varphi}_1,\ldots,\overline{\varphi}_n)))$$ 
    By the same proof steps as for filterable quantifiers, we have that $\delta_I \preceq |\mu_I|^*_\tau(\sem{\prod_F \cM}(\tau.\overline{\Qf}(\overline{\varphi}_1,\ldots,\overline{\varphi}_n)))$, and then $\delta_I \not\preceq |\mu_I|^*_\tau(\sem{\prod_F \cM}(\tau.\Qf(\varphi_1,\ldots,\varphi_n)))$.
    \item $(\Leftarrow)$ Let us suppose that $\delta_I \not\preceq |\mu_I|^*_\tau(\sem{\prod_F \cM}(\tau.\cQ f(\varphi_1,\ldots,\varphi_n)))$. By duality, there is \(S \subseteq \{1,\ldots,n\}\) such that
    $$\delta_I \preceq |\mu_I|^*_\tau(\sem{\prod_F \cM}(\tau.\overline{\Qf}(\overline{\varphi}_1,\ldots,\overline{\varphi}_n)))$$  
    Therefore, we have that 
    $$\{i \in I \mid \delta_I \not\preceq |p_{I,i}|^*_\tau(\sem{\cM_i}(\tau.\cQ f(\varphi_1,\ldots,\varphi_n))\} \in F$$
    and then
    $$\{i \in I \mid \delta_I \preceq |p_{I,i}|^*_\tau(\sem{\cM_i}(\tau.\cQ f(\varphi_1,\ldots,\varphi_n))\} \notin F$$
\end{itemize}

\subsection{Compactness theorem}
\label{subsec:los:compactness}

In standard model theory, a direct application of \Los's result is the compactness theorem. However, this application only holds for a subset of formulas, namely, sentences. In FOL, sentences are formulas without free variables, i.e., formulas over the empty context $[]$. For FOL in an elementary topos $\cC$ (see Section~\ref{fol-semanticalsystem}), for a given $\Sigma$-model $\cM$, $\Sub(|\cM|([]))$ is the Heyting algebra with only two elements the upper and lower bounds (where \([]\) is the empty context ensuring that \(|\cM|([])\) is the terminal object of \(\cC\)). This characterization naturally leads to the following definition of sentences.

\begin{definition}[Sentence]
    A formula \(\sigma.\varphi \in \cF_\cL\) is a {\bf sentence} if for all models \(\cM \in |Mod|\), $\sem{\cM}(\sigma.\varphi)$ is either $\bot_{|\cM|(\sigma)}$ or $\top_{|\cM|(\sigma)}$. 
\end{definition}

\begin{proposition}
\label{cor:Los}
If \(\sigma.\varphi \in \cF_\cL\) is a sentence, then:
\[
    \prod_F \cM \models \sigma.\varphi \; \mbox{iff} \; \{i \in I \mid \cM_i \models \sigma.\varphi\} \in F.
\]
\end{proposition}

\begin{proof}
Let us suppose that $\prod_F \cM \models \sigma.\varphi$. This means that for all generators $\delta_I$ in $l_{|\prod_I \cM_i|(\sigma)}$, $\delta_I \preceq |\mu_I|^*_\sigma(\prod_F \cM \models_\iota \sigma.\varphi))$. By Theorem~\ref{th:Los}, this means that $J = \{i \in I \mid \delta_I \preceq |p_{I,i}|^*_\sigma(\sem{\cM_i}(\sigma.\varphi))\} \in F$, from which we can conclude for every $j \in J$ that $\cM_j \models \sigma.\varphi$.

\medskip
Let us suppose that $J = \{i \in I \mid \cM_i \models \sigma.\varphi\} \in F$. This means that for every $j \in J$, and every $\delta_I \in l_{|\prod_I \cM_i|(\sigma)}$ that $\delta_I \preceq |p_{I,j}|^*_\sigma(\sem{\cM_j}(\sigma.\varphi))$. By Theorem~\ref{th:Los}, we have that $\delta_I \preceq |\mu_I|^*_\sigma(\sem{\prod_F \cM}(\sigma.\varphi))$, and then $\prod_F \models \sigma.\varphi$.
\end{proof}

The compactness result follows via this definition and the classical proof using ultraproducts and \Los's theorem.

\begin{theorem}[Compactness]
\label{th:compactness}
A set of sentences $T$ has a model if and only if every finite subset of $T$ has a model.
\end{theorem}

\begin{proof}
    ($\Rightarrow$) Let $\cM$ be a model of $T$. Since $\cM \models T$, then $\cM$ satisfies every finite subset of $T$.

    \medskip
    \noindent
    ($\Leftarrow$)
    Suppose that every finite subset of $T$ admits a model. Let $I$ denote the collection of all finite subsets of $T$. For each $i \in I$, let $\cM_i$ be a model such that  $\cM_i\models i$. Define $i^* = \{j \in I \mid i \subseteq j\}$ for every $i \in I$, and $I^* = \{i^* \mid i \in I\}$. It is straightforward to show that $I^*$ has the finite intersection property. By Zorn's lemma, there exists an ultrafilter $F$ such that $I^* \subseteq F$. Consider the filtered product $\prod_F \cM$ of the family of models $(\cM_i)_{i \in I}$. Choose a sentence $\sigma.\varphi \in T$. Since $\{\sigma.\varphi\} \in I$, it follows that $\cM_i \models \sigma.\varphi$ for all $i \in I$ with $\sigma.\varphi \in i$. Hence, we obtain that
    $$\{\sigma.\varphi\}^* \subseteq \{j \in I \mid \cM_j \models \sigma.\varphi\}$$
    As $\{\sigma.\varphi\}^* \in F$, we have that $J = \{j \in I \mid \cM_j\models \sigma.\varphi\} \in F$. Since $\sigma.\varphi$ is a sentence, according to Proposition~\ref{cor:Los}, we deduce that $\prod_F \cM \models \sigma.\varphi$.
\end{proof} %
\section{Conclusion}

We studied the mathematical construction of ultraproducts within the framework of an abstraction of categorical logic.
Ultraproducts have proven their significance in universal algebra and mathematical logic, particularly in model theory.
We explored how to adapt \Los's theorem, also known as the fundamental theorem of ultraproducts, which was initially established for first-order logic with set models, to our abstract categorical logic, thereby making it independent of any specific quantifier.

Due to the intrinsic abstract nature of our logical formalism, we have imposed some technical yet natural conditions on our logical system. These conditions are the following:

\begin{itemize}
    \item First, the underlying Heyting algebras of the prop-categories should be sup-generated in a finite way (first and last condition of Definition~\ref{def:filteredsemsyst}). This first condition can be related to the construction of locally finitely presented categories and solve the difficulty of filters only being closed under finite intersections.
    \item Secondly, the prop-categories should be covered to preserve disjunctive formulas along model ultraproducts.
    \item Obviously, the category of models should have filtered products.
    \item Finally, quantifiers should be filterable (which can be interpreted as the inverse implication of isotonicity) to be able to consider the result independently of the quantifiers. This condition has been extended to pullback functors to deal with formulas of the form $f(\varphi)$.
\end{itemize}

While explaining the motivation behind each condition, we have also shown their relevance through a series of examples from various logical formalisms. As a direct application, we have also derived an abstract compactness result, leveraging a semantical definition of sentences.

\medskip
For future work, we plan to study the generalization of other model theoretical results within the framework of abstract categorical results. Additionally, we defined a complete formal system akin to sequent calculus in~\cite{AB23} for which we plan to explore its proof-theoretic aspects, for instance, to obtain equivalents of theorems such as Barr's theorem~\cite{barr_toposes_1974} which states that within the framework of geometric logic, if a geometric sentence is deducible from a geometric theory in classical logic, with the axiom of choice, then it is also deducible from it intuitionistically~\cite{Wraith78}. To get around the fact that the proof of Barr's theorem is non-constructive, we could also see how to adapt the different methods developed within FOL and HOL, which typically consist of transforming classical proofs into intuitionistic ones by adding double negations in suitable places.

\section*{Declaration of competing interest}

This work was partly funded by Isabelle Bloch's Chair in Artificial Intelligence (Sorbonne Université and SCAI) and by the Deutsche Forschungsgemeinschaft (DFG, German Research Foundation) -- CRC 1608 -- 501798263.

\section*{Data availability}
No data was used for the research described in the article. 
\bibliographystyle{elsarticle-num-names} 
\bibliography{cleaned-biblio}

\appendix

\section{Logic and internal language in topos}
\label{sec:AppB}

An interesting feature of toposes
is that we can reason on objects and morphisms of a topos ``as if they were sets and functions"~\cite{Johnstone02,Law72}. The reason is that we can do logic in toposes. Indeed, we can define logical connectives in toposes. Here, we recall the definition of propositional connectives 
$\{\wedge,\vee,\neg,\Rightarrow\}$ and of constants $true, false$. %
\begin{itemize}
    \item By definition of subobject classifiers, we have a monomorphism $true : \mathbb{1} \rightarrowtail \Omega$, and then we also have a morphism $(true,true) : \mathbb{1} \rightarrowtail \Omega \times \Omega$ which is also a monomorphism. So, by the subobject classifier definition, $\wedge : \Omega \times \Omega \to \Omega$ is its characteristic morphism. 
    \item {$\vee : \Omega \times \Omega \to \Omega$ classifies the image of the morphism $[(true,Id_\Omega),(Id_\Omega,true)] : \Omega +\Omega \to \Omega \times \Omega$,} {where $+$ denotes the co-product.}
    \item the morphism $\Rightarrow : \Omega \times \Omega \to \Omega$ is the characteristic morphism of $\preceq \rightarrowtail \Omega \times \Omega$ where $\preceq$ is the equalizer of $\wedge$ and the projection on the first argument $p_1 : \Omega \times \Omega \to \Omega$. 
    \item Finally, the unique morphism $\emptyset \rightarrowtail \mathbb{1}$ is a monomorphism. Let us denote by $false : \mathbb{1} \to \Omega$ its characteristic morphism. Then, $\neg : \Omega \to \Omega$ is defined as the composite $\Rightarrow \circ \, (Id_\Omega \times false)$. 
\end{itemize}
Consequently the power object $\Omega = P\, \mathbb{1}$ is an internal Heyting algebra\footnote{An {\em internal Heyting algebra} in a topos is an internal lattice $L$, that is equipped with two morphisms $\wedge,\vee : L \times L \to L$ such that the diagrams expressing the standard laws for $\wedge$ and $\vee$ commute, and with top and bottom which are morphisms $\bot,\top: \mathbb{1} \to L$ such that $\wedge\circ (Id_L \times \top) = Id_L$ and $\vee \circ (Id_L \times \bot) = Id_L$, together with an additional morphism $\Rightarrow : L \times L \to L$ which satisfies the diagrams given by the
identities:
\begin{itemize}
    \item $x \Rightarrow x = \top$
    \item $x \wedge (x \Rightarrow y) = x \wedge y$ and $y \wedge (x \Rightarrow y) = y$
    \item $x \Rightarrow (y \wedge z) = (x \Rightarrow y) \wedge (x \Rightarrow z)$
\end{itemize}
}
and then the logic is intuitionistic. Actually, through the bijection $\Sub(X \times Y) \simeq \Hom_\cC(X,PY)$, for every object $X$ in a topos $\cC$, $PX$ is an internal Heyting algebra. We can then define a partial order $\preceq_X$ 
as an object of $\cC$ such that $\preceq_X$ is the equalizer of $\wedge : PX \times PX \to PX$ and $p_1 : PX \times PX \to PX$ where $p_1$ is the projection on the first argument of couples. %

For every topos $\cC$, we can define an internal language $\cL_\cC$ composed of types defined by the objects of $\cC$, from which we can define terms as follows:

\begin{itemize}
\item $true:X$;
\item $x:X$ where $x$ is a variable and $X$ is a type;
\item $f(t):Y$ where $f : X \to Y$ is a morphism of $\cC$ and $t:X$ is a term;
\item $<t_1,\ldots,t_n>: X_1 \times \ldots \times X_n$ if for every $i$, $1 \leq i \leq n$, $t_i:X_i$ is a term;
\item $(t)_i:X_i$ if $t:X_1 \times \ldots \times X_n$ is a term;
\item $\{x:X \mid \alpha\}:PX$ if $\alpha:\Omega$ is a term;
\item $\sigma = \tau:\Omega$ if $\sigma$ and $\tau$ are terms of the same type;
\item $\sigma \in_X \tau:\Omega$ if $\sigma:X$ and $\tau:PX$ are terms;
\item $\sigma \preceq_X \tau:\Omega$ if $\sigma,\tau:PX$ are terms;
\item $\varphi~@~\psi:\Omega$ if $\varphi:\Omega$ and $\psi:\Omega$ are terms with $@ \in \{\wedge,\vee,\Rightarrow\}$;
\item $\neg \varphi:\Omega$ if $\varphi:\Omega$ is a term;
\item $Q x.\,\varphi:\Omega$ if $x:X$ and $\varphi:\Omega$ are terms and $Q \in \{\forall,\exists\}$.
\end{itemize}
Terms of type $\Omega$ are called {\em formulas}.

Semantics of terms will depend on their type. Hence, semantics of terms of type $X \neq \Omega$ will be defined by morphisms, and terms of type $\Omega$ will be interpreted as subobjects. 

We say that a sequence of variables $\vec{x} = (x_1,\ldots,x_n)$ is a {\em suitable context} for a term or a formula if each free variable of this term or this formula occurs in $\vec{x}$. Let us denote by $X_{\vec{x}}$ the product $X_1 \times \ldots \times X_n$ when $\vec{x} = (x_1,\ldots,x_n)$ and each $x_i : X_i$. Then the {\em semantics} of $t:X$ in the context $\vec{x}$, denoted by $\sem{t}_{\vec{x}}$, is a morphism from $X_{\vec{x}}$ to $X$.  It is defined recursively on the structure of $t$ as follows:

\begin{itemize}
\item $\sem{x_i:X_i}_{\vec{x}} = p_i$ where $p_i : X_{\vec{x}} \to X_i$ is the obvious projection on the $i^{th}$ argument;
\item $\sem{f(t)}_{\vec{x}} = f \circ \sem{t}_{\vec{x}}$;
\item $\sem{<t_1,\ldots,t_n>}_{\vec{x}} = (\sem{t_1}_{\vec{x}},\ldots,\sem{t_n}_{\vec{x}})$;
\item $\sem{(t)_i}_{\vec{x}} = p_i \circ \sem{t}_{\vec{x}}$ where $p_i$ is the projection on the $i^{th}$ argument of the tuple;
\item $\sem{\{x:X \mid \alpha\}}_{\vec{x}}$ is the unique morphism $r : X_{\vec{x}} \to PX$ making the diagram below a pullback square
\[\begin{tikzcd}[column sep = 5em]
	R & {\in_X} \\
	{X \times X_{\vec{x}}} & {X \times PX}
	\arrow[from=1-1, to=1-2]
	\arrow[tail, from=1-2, to=2-2]
	\arrow["{\sem{\alpha}_{(x,\vec{x})}}"', tail, from=1-1, to=2-1]
	\arrow["{Id_X \times r}"', from=2-1, to=2-2]
\end{tikzcd}\]
\end{itemize}

The semantics of a formula $\varphi:\Omega$ in the context $\vec{x}$, denoted by $\sem{\varphi}_{\vec{x}}$, is interpreted as a subobject of $\Sub(X_{\vec{x}})$ and is recursively defined as follows:
 
\begin{itemize}
\item $\sem{true}_{\vec{x}} = Id_{X_{\vec{x}}}$; 
\item when $\varphi = \sigma = \tau$, then $\sem{\varphi}_{\vec{x}}$ equalizes $\sem{\sigma}_{\vec{x}}$ and $\sem{\tau}_{\vec{x}}$;
\item when $\varphi = \sigma \in_X \tau$, then $\sem{\varphi}_{\vec{x}} : R \rightarrowtail X_{\vec{x}}$ where $R$ is the pullback of the diagram
\[\begin{tikzcd}[column sep = 5em]
	R & {\in_X} \\
	{X_{\vec{x}}} & {X \times PX}
	\arrow[from=1-1, to=1-2]
	\arrow[tail, from=1-2, to=2-2]
	\arrow["{\sem{\varphi}_{\vec{x}}}"', tail, from=1-1, to=2-1]
	\arrow["{\sem{\sigma}_{\vec{x}} \times \sem{\tau}_{\vec{x}}}"', from=2-1, to=2-2]
\end{tikzcd}\]
\item if $\varphi = \sigma \preceq_X \tau$, then $\sem{\varphi}_{\vec{x}} : R \rightarrowtail X_{\vec{x}}$  where $R$ is the pullback of the diagram 
\[\begin{tikzcd}[column sep = 5em]
	R & {\preceq_X} \\
	{X_{\vec{x}}} & {PX \times PX}
	\arrow[from=1-1, to=1-2]
	\arrow[tail, from=1-2, to=2-2]
	\arrow["{\sem{\varphi}_{\vec{x}}}"', tail, from=1-1, to=2-1]
	\arrow["{\sem{\sigma}_{\vec{x}} \times \sem{\tau}_{\vec{x}}}"', from=2-1, to=2-2]
\end{tikzcd}\]
\item if $\varphi = \varphi_1~@~\varphi_2$, then $\sem{\varphi}_{\vec{x}} = \sem{\varphi_1}_{\vec{x}}~@~\sem{\varphi_2}_{x_t}$ where $@$ is the operator in $\{\wedge,\vee,\Rightarrow\}$ in the Heyting algebra $\Sub(X_{\vec{x}})$;
\item $\sem{\neg \varphi}_{\vec{x}} = \neg_{X_{\vec{x}}} (\sem{\varphi}_{\vec{x}})$ where $\neg_{X_{\vec{x}}}(\sem{\varphi}_{\vec{x}})$ is the pseudo-complement of $\sem{\varphi}_{\vec{x}}$ in $\Sub(X_{\vec{x}})$;
\item $\sem{\forall x.\varphi}_{\vec{x}} = \forall_p(\sem{\varphi}_{(\vec{x},x)})$ where $p : X_{\vec{x}} \times X \to X_{\vec{x}}$ is the projection, and $\forall_p$ is the right adjoint to the pullback functor $p^* : \Sub(X_{\vec{x}}) \to \Sub(X_{\vec{x}} \times X)$ when the Heyting algebras $\Sub(X_{\vec{x}})$ and $\Sub(X_{\vec{x}} \times X)$ are regarded as categories. 
\item $\sem{\exists x.\varphi}_{\vec{x}}$ is the image of $p \circ \sem{\varphi}_{(\vec{x},x)}$ where $p$ is the same projection as above.
\end{itemize}
Equivalently, semantics of any formula $\varphi:\Omega$ could be defined by a morphism from $X_{\vec{x}}$ to $\Omega$, by interpreting $\varphi$ as the classifying morphism of $\sem{\varphi}_{\vec{x}}$. 

We write $\cC \models_{\vec{x}} \varphi$ if $\sem{\varphi}_{\vec{x}} = Id_{X_{\vec{x}}}$ ($Id_{X_{\vec{x}}}$ is the top element in $\Sub(X_{\vec{x}})$).

\end{document}